%% file: main.tex
\documentclass[12pt,draftclsnofoot,onecolumn]{IEEEtran}
\usepackage{xcolor}
\usepackage{balance}
\usepackage{stfloats}
\usepackage{cite}
\usepackage{amsmath,amssymb,amsfonts}
\usepackage{graphicx}
\usepackage{textcomp}
\usepackage{acronym}
\usepackage{xcolor}
\usepackage{tikz} % for tikz
\usepackage[utf8]{inputenc}
\usepackage{pgfplots} 
\usepackage{pgfgantt}
\usepackage{pdflscape}
\usepackage{changes}
\usepackage{comment}
\usepackage{subfigure}
\usepackage{mathtools,algpseudocode,algorithm,MnSymbol}
\usepackage{geometry}
\DeclareMathOperator*{\argmax}{arg\,max}
\DeclareMathOperator*{\argmin}{arg\,min}

\usepackage{pgfplots}
  \pgfplotsset{compat=newest}
  \usetikzlibrary{plotmarks}
  \usetikzlibrary{arrows.meta}
  \usepgfplotslibrary{patchplots}
  \usepackage{grffile}
  \usepackage{amsmath}

\pgfplotsset{compat=newest} 
\pgfplotsset{plot coordinates/math parser=false} % end of tikz

\input{acronyms.tex}

\begin{document}

\bibliographystyle{IEEEtran}
\bstctlcite{IEEEexample:BSTcontrol}
%\title{An End-to-end 5G SLAM Framework with a Low-complexity Channel Estimator}
\title{Experimental Validation of Single BS 5G mmWave Positioning and Mapping for Intelligent Transport}

\author{Yu Ge,~\IEEEmembership{Student~Member,~IEEE,} Hedieh Khosravi,~\IEEEmembership{Student~Member,~IEEE,} \\ Fan Jiang,~\IEEEmembership{Member,~IEEE,} Hui Chen,~\IEEEmembership{Member,~IEEE,} Simon Lindberg, Peter Hammarberg,  Hyowon Kim,~\IEEEmembership{Member,~IEEE,} Oliver Brunneg\aa rd,
 Olof Eriksson,  
Bengt-Erik Olsson, 
\\Fredrik Tufvesson,~\IEEEmembership{Fellow,~IEEE,} 
Lennart Svensson,~\IEEEmembership{Senior~Member,~IEEE,}     \\ 
Henk Wymeersch,~\IEEEmembership{Senior~Member,~IEEE}
\thanks{Yu Ge, Fan Jiang, Hui Chen, Hyowon Kim, Lennart Svensson and Henk Wymeersch are with the Department of Electrical Engineering, Chalmers University of Technology, Gothenburg, Sweden. Emails: 
\{yuge, fan.jiang, hui.chen, hyowon, lennart.svensson, henkw\}@chalmers.se.}
\thanks{Hedieh Khosravi and Fredrik Tufvesson are with the Department of Electrical and Information Technology, Lund University, Lund, Sweden. Emails: \{hedieh.khosravi,fredrik.tufvesson\}@eit.lth.se.}
\thanks{Simon Lindberg is with Qamcom, Gothenburg, Sweden. Email: simon.lindberg@qamcom.se.}
\thanks{Peter Hammarberg and Bengt-Erik Olsson are with Ericsson Research, Gothenburg, Sweden. Email: \{peter.hammarberg, bengt-erik.olsson\}@ericsson.com.}
\thanks{Oliver Brunneg\aa rd and Olof Eriksson are with Veoneer, V\aa rg\aa rda, Sweden. Email: \{oliver.brunnegard, olof.eriksson\}@veoneer.com.}
\thanks{This work was partially supported by the Vinnova 5GPOS project under grant 2019-03085, by the Swedish Research Council under grant 2018-03705, and by the Wallenberg AI, Autonomous Systems and Software Program (WASP) funded by Knut and Alice Wallenberg Foundation.}
}


\maketitle
\pagestyle{empty}
\thispagestyle{empty}
\begin{abstract}
Positioning with 5G signals generally requires connection to several \acp{bs}, which makes positioning more demanding in terms of infrastructure than communications. To address this issue, there have been several theoretical studies on single \ac{bs} positioning, leveraging high-resolution angle and delay estimation and multipath exploitation possibilities at mmWave frequencies. This paper presents the first realistic experimental validation of such studies, involving a commercial 5G mmWave \ac{bs} and a \ac{ue} development kit mounted on  a test vehicle. We present the relevant signal models, signal processing methods (including channel parameter estimation and position estimation), and validate these based on real data collected in an outdoor science park environment. Our results indicate that  positioning is possible, but 
the performance with a single \ac{bs} is limited by the knowledge of the position and orientation of the infrastructure and the multipath visibility and diversity. 
\end{abstract}

\vskip0.5\baselineskip

% \fi
% \IEEEpeerreviewmaketitle

%This paper will mainly focus on technical details of the component of the systems and the algorithms to realize positioning and mapping.

%Target journal: IEEE Journal on Selected Areas in Communications (5G/6G Precise Positioning on Cooperative Intelligent Transportation Systems (C-ITS) and Connected Automated Vehicles (CAV))

%Deadline: December 1st

\section{Introduction}
%1.5 pages
\acresetall

5G and Beyond 5G communication systems are expected to play an important role in intelligent transportation \cite{bartoletti2021positioning}, complementing on-board sensors such as radar, lidar, and \ac{gnss}, by providing an absolute position estimate, even in harsh urban and indoor environments. 
To realize 5G/B5G positioning, standardization efforts in 3GPP and ETSI as well as different means of technical enhancements are provided to support a variety of commercial use cases with different performance requirements~\cite{tr:38845-3gpp21,tr:38859-3gpp22}.
%Standardization efforts in 3GPP and ETSI focus on supporting a variety of commercial use cases with different positioning performance requirements \cite{tr:38845-3gpp21}, by means of technical enhancements \cite{tr:38859-3gpp22}. 
The main technical enabler for accurate positioning in 5G is the ability to operate at mmWave frequencies above 24 GHz, which brings a number of concrete benefits \cite{wymeersch20175g,dwivedi2021positioning,tr:38855-3gpp19}. First of all, large bandwidths in the order of 400 MHz are available at mmWave, providing high delay resolution. Second, within a fixed footprint, more  antenna elements can be deployed in mmWave compared to sub-6 GHz, at both \ac{bs} and \ac{ue} sides. This provides high angular resolution and the possibility of \ac{ue} orientation estimation. Third, the propagation channel at mmWave tends to be more geometric and less random than at sub-6 GHz carriers, making practical and scalable model-based positioning algorithms 
possible. %, where machine learning methods provide reasonable positioning performance  \cite{butt2020rf}.

An important limitation of 5G mmWave positioning is the high demand in terms of infrastructure. 5G mmWave communication links typically provide coverage of no more than a few hundred meters, which is further compounded by the fact that conventional \ac{TDOA} positioning  requires simultaneous connections to at least four synchronized \acp{bs} \cite{keating2019overview} and typically many more for specific scenarios like highways \cite{del2016feasibility}. Multi-\ac{bs} solutions also require optimized deployments \cite{posluk20215g} and accurate calibration processes to obtain anchor positions. 

To reduce the infrastructure and deployment cost, the community has made a concerted effort to devise alternative approaches whereby 3D positioning of a \ac{ue} can be accomplished with a single \ac{bs}. These can be characterized in roughly four categories: (i) single-\ac{bs} positioning using \ac{rtt} and \ac{AOD} (in downlink) or \ac{AOA} (in uplink) \cite{guo2022performance,blanco2019performance,kanhere2018position}; (ii) single-\ac{bs} positioning using information fusion \cite{mostafavi2019vehicular,mostafavi2020vehicular} (e.g., from external sensors); (iii) single-\ac{bs} positioning using \acp{RIS} \cite{Henk_RISSLAM_VTM2020,zhang2022toward,fascista2021ris,keykhosravi2021siso}; (iv) single-\ac{bs} positioning based on passive multipath information \cite{witrisal2016high,Shahmansoori18,wen20195g,NLOSforPositionOrientationEstimation--R.Mendrzik_H.Wymeersch_G.Bauch,ge2022computationally,ge2022mmwave,sun2021comparative,wymeersch20185g,liu2017prospective}. 
In particular, turning multipath from foe to friend in (iv) is attractive, as it comes at no cost in terms of time-frequency resources (in contrast to approach (i)), it does not require external sensors to provide a position fix (in contrast to approach (ii)), and it does not require additional infrastructure (in contrast to approach (iii)). 

\begin{figure}
    \centering
\includegraphics[width=0.6\columnwidth]{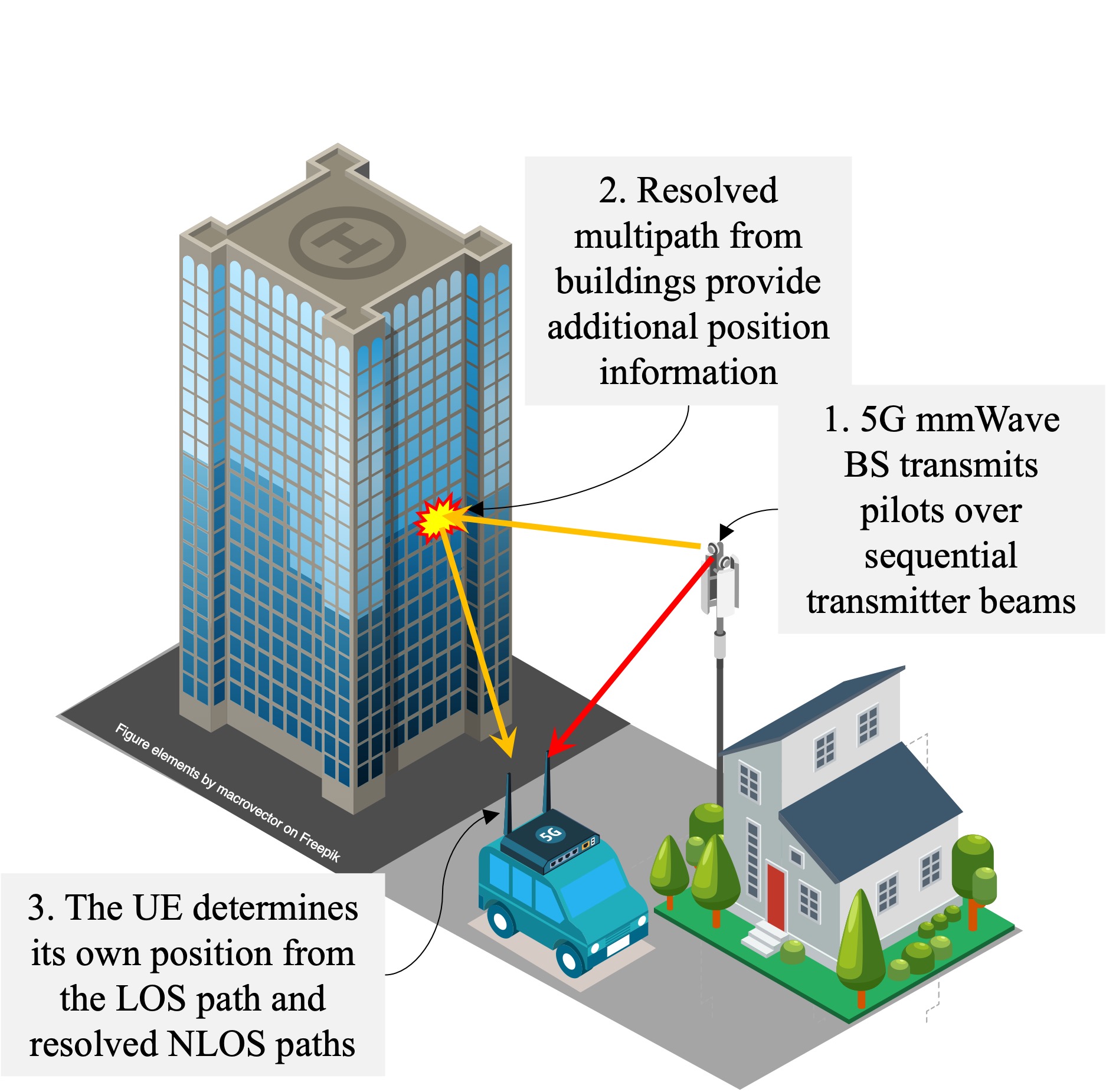}
    \caption{Exemplifying scenario of single-BS positioning, harnessing delay, \ac{AOD}, and \ac{AOA} information from the \acf{los} and resolved \acf{nlos} paths. } \vspace{-5mm}
    \label{fig:overview}
\end{figure}

Despite significant theoretical progress, only a limited subset of the above approaches has been shown to operate in practice \cite{yammine2021experimental,mata2020preliminary,witrisal2016high,ge2022experimental}. 
In \cite{yammine2021experimental}, a 5G FR2 experimental setup for indoor positioning based on \ac{TOA} was demonstrated, and an analysis of the impact of beam pairing was conducted. In \cite{mata2020preliminary}, results on indoor and outdoor trials in 5G FR1 were reported, yielding ranging errors from sub-meter indoors to several hundreds of meters outdoors, indicating that 5G measurements should be integrated with other sensors (e.g., \ac{gnss}) to meet requirements for \ac{AD}
 and \ac{ADAS} applications. A demonstration of combined \ac{TOA} and \ac{AOA} LTE positioning was demonstrated in \cite{blanco2019performance}, while \cite{kanhere2018position} evaluated the combination of  \ac{RSS} and \ac{AOA} in 5G mmWave. 
The concept of multipath exploitation was validated in \cite{witrisal2016high} for an indoor scenario following the IEEE 802.11ad standard (with 63 GHz carrier and 2 GHz bandwidth). 
Apart from our preliminary results \cite{ge2022experimental} on  \ac{bs} calibration, there have been no reported demonstrations of 5G mmWave vehicular positioning using commercial devices.

%3GPP \cite{ts:38211-3gpp20,tr:38855-3gpp19}.

%Fusion paper possibility of single \ac{bs} positioning \cite{ge2022experimental}.

%Single Base Station ToA-AoA Positioning in an LTE Testbed \cite{blanco2019performance}, even though it is not 5G, it is radio and also has experiments.

%A magazine \cite{liu2017prospective}.

%5G ranging measurements fuse with GNSS, has experiments\cite{yin2018gnss,wang2022simulation}

%This also might be useful \cite{kanhere2018position}

%Studies reporting experimental validation of 5G mmWave positioning have been limited \cite{yammine2021experimental,mata2020preliminary,ge2022experimental}, and most existing works rely on simulations \cite{dwivedi2021positioning,keating2019overview}.

In this paper, we present the results of a unique demonstration of single BS 5G mmWave positioning of a vehicle using commercial hardware and a real vehicle. Our demonstration sheds new light on the capabilities of 5G to support intelligent transportation applications as well as several limitations that should be addressed in future standards. The main contributions are summarized as follows. 
 \begin{itemize}
        \item \textbf{Single-BS positioning system overview:} We introduce the hardware specification for the 5G single base-station positioning system, including the commercial \ac{bs}, the \ac{ue}, the ground-truth system, and the outdoor operating environment. Particular emphasis is placed on the UE device mounting, UE and BS array configurations, and utilized beam patterns. %We introduce the hardware specification for the 5G single base-station positioning systems, including the commercial BS, the \ac{ue}, and the ground-truth system.
        \item \textbf{End-to-end signal processing description:} We provide technical algorithms for processing the 5G mmWave measurements for positioning purposes, including channel parameter estimation, BS calibration, and positioning and mapping algorithms.  In particular, we provide low-complexity algorithms for single-BS \ac{los}-only positioning using RTT and AOD, as well as mixed LOS and \ac{nlos} positioning with multipath exploitation. Moreover, a simple environment mapping algorithm is presented, given the estimated UE position.%We provide technical algorithms for processing the real 5G measurements for positioning purposes, including channel estimation, BS calibration, and positioning and mapping algorithms.
        \item \textbf{Performance evaluation with real data:} We show positioning and mapping results using real 5G measurements and analyze the performance gap via different combinations of real and simulated measurements. We found that the combination of RTT, AOD and AOA provides around 1.71 m positioning accuracy (90\% of cases), but can be further improved with super-resolution methods. Positioning with multipath aiding and \ac{TDOA} turns out to be severely limited by the propagation environment and yielded errors in excess of 10 meters. Correspondingly, mapping performance is shown to be much better when RTT is available.
        %\item \textbf{LOS and mixed LOS \& NLOS positioning and mapping:}  We provide algorithms for single-BS LOS-only positioning using RTT and AOD, as well as mixed LOS and NLOS positioning with multipath exploitation. 
        %\item \textbf{Performance evaluation with real data:} We show positioning and mapping results using real 5G measurements and analyze the performance gap via different combinations of real and simulated measurements. %, and provide outlook for future, using generated map (if necessary), different levels of clock bias noise, different channel estimation.
    \end{itemize}
%\begin{itemize}
%\item What, and why it is important, its benefits

%\item Background, 5G provides new opportunities

%\item Literature reviews on previous works, pros \& cons 

%\item Challenges

%\item Contributions
%    \begin{itemize}
 %       \item Introduce the hardware details about the 5G single base-station positioning (potential mapping) systems, including BS, user, GPS;
  %      \item Provide necessary technical algorithms on processing the real 5G measurements to achieve positioning purpose, including channel estimation, BS calibration, and MAP estimator using only LOS for positioning;
   %     \item Provide algorithms on utilizing NLOS and real-time positioning (SLAM), including offline LOS + NLOS positioning and mapping, and SLAM filter;
    %    \item Show positioning and mapping using real 5G measurement;
     %   \item Analyze the performance gap via using different combination of real measurements and simulated measurement, and provide outlook for future, using generated map (if necessary), different levels of clock bias noise, different channel estimation.
    %\end{itemize}

%\end{itemize}

The remainder of this paper is structured as follows. The system models are described in Section \ref{models}. The system components are then introduced in Section \ref{sec:Components}. The channel estimation, %are displayed in Section \ref{Section:channel_estimation}, and
positioning and mapping algorithms are derived in Section \ref{sec:localization_mapping}. Tests and results are presented in Section \ref{results}, followed by our conclusions in Section \ref{Conclusions}.

\subsubsection*{Notations}Scalars (e.g., $x$) are denoted in italic, vectors (e.g., $\boldsymbol{x}$) are denoted in bold lower-case letters, and matrices (e.g., $\boldsymbol{X}$) are denoted in bold capital letters. The transpose is denoted by $(\cdot)^{\mathsf{T}}$, the Hermitian transpose is denoted by $(\cdot)^{\mathsf{H}}$, the conjugate is denoted by $(\cdot)^{*}$, the Euclidean norm is denoted by $\lVert \cdot \rVert$, the $n$-th component of a vector is denoted by $[\cdot]_{n}$, while $[\cdot]_{n:n'}$ extracts components $n$ until $n'>n$, 
a Gaussian distribution with mean $\boldsymbol{u}$ and covariance $\boldsymbol{\Sigma}$, evaluated in value $\boldsymbol{x}$, is denoted by $\mathcal{N}(\boldsymbol{x};\boldsymbol{u},\boldsymbol{\Sigma})$.

\section{System models}\label{models}
%{1.5 pages}
In this section, we introduce the state models of the \ac{ue}, \ac{bs} and landmark in a typical 5G mmWave downlink scenario, where a single \ac{ue} mounted on a vehicle, a single \ac{bs} mounted on a tower,  and a few others landmarks, e.g., buildings or walls, are included. Moreover, the received signal model and the measurement model are also provided.

\subsection{State Models}
We use the global reference coordinate system as the reference in this paper.
The \ac{bs} is fixed and equipped with a \ac{URA}, where its center is located at $\boldsymbol{p}_\text{BS} = [x_{\text{BS}}, y_{\text{BS}}, z_{\text{BS}}]^{\mathsf{T}}$. It has a rotation of $\boldsymbol{\psi}_{\text{BS}}=[\alpha_\text{BS}, \beta_\text{BS}, \gamma_\text{BS}]^{\mathsf{T}}$, denoting the roll, pitch and yaw in the respective order, 
which describes the orientation of the transmitter \ac{URA} with respect to the reference coordinate system \cite{blanco2010tutorial}.
%which describes the direction of the transmitter \ac{URA} and shows how to rotate the reference coordinate system to the local coordinate system of the transmitter \ac{URA} \cite{blanco2010tutorial}.
Therefore, the state of the \ac{bs} can be modeled as $\boldsymbol{s}_\text{BS}=[\boldsymbol{p}_{\text{BS}}^{\mathsf{T}}, \boldsymbol{\psi}_{\text{BS}}^{\mathsf{T}}]^{\mathsf{T}}$, which is usually known a priori.

The \ac{ue} moves on the ground over time. It is also equipped with an \ac{URA}, and has the state $\boldsymbol{s}_{\text{UE},k}=[\boldsymbol{p}_{\text{UE},k}^{\mathsf{T}}, \boldsymbol{\psi}_{\text{UE},k}^{\mathsf{T}},b_{k}]^{\mathsf{T}}$ at time $k$, with
$\boldsymbol{p}_{\text{UE},k}=[x_{\text{UE},k}, y_{\text{UE},k},z_{\text{UE},k}]^{\mathsf{T}}$ denoting the position of the center of the \ac{URA} at the receiver side, $\boldsymbol{\psi}_{\text{UE},k}=[\alpha_{\text{UE},k}, \beta_{\text{UE},k}, \gamma_{\text{UE},k}]^{\mathsf{T}}$ representing the direction of the receiver, and $b_{k}$ denoting the clock bias caused by the imperfect synchronization between the transmitter and the receiver. 

Apart from the \ac{bs}, there are a few other landmarks in the scenario, like buildings and walls, and their facades can reflect and/or diffuse downlink signals to the receiver, which results in \ac{nlos} paths. Therefore, we describe the environment with some effective surfaces and parameterize the point where the signal hit the effective surface as an \ac{IP} with location $\boldsymbol{p}_{\mathrm{IP}}$. \acp{IP} of those \ac{nlos} paths which are reflected or diffused by the same source are on the same surface by definition, so that \acp{IP} reveal where surfaces exist.  %and we can estimate surfaces from \acp{IP}.

%Apart from the \ac{bs}, there a few other landmarks in the scenario, like some buildings or walls, and their facades can reflect or/and diffuse downlink signals to the receiver. Therefore, we describe the environment with some effective surfaces. Each effective surface can be parameterized by a fixed \ac{VA} with location $\boldsymbol{p}_{\mathrm{VA}}$, which is the reflection of the  \ac{bs} with respect to the effective surface{\cite{Rappaport_6G100GHz_Access2019,palacios2019single}}
% kim20205g}:
%\begin{align}
    %\boldsymbol{p}_{\mathrm{VA}}=(\boldsymbol{I}-2\boldsymbol{\nu}\boldsymbol{\nu}^{\mathsf{T}}) \boldsymbol{p}_{\mathrm{BS}}+2\boldsymbol{\mu}^{\mathsf{T}}\boldsymbol{\nu} \boldsymbol{\nu},\label{VA}
%\end{align}
%where $\boldsymbol{\nu}$ is the normal to the reflecting surface and $\boldsymbol{\mu}$ is an arbitrary point on the surface. From \eqref{VA}, we could obverse that the \ac{VA} does not depend on the \ac{ue} state, so that although the \acp{IP} where the signals hit the surface change with \ac{ue} moving around, the VA remains static. Moreover, the \ac{VA} is surface-specific, and when the \ac{VA} is given, the surface can be determined.

\subsection{Signal Model}
%\begin{itemize}
%    \item Received OFDM signal model (measurements as close as)
%    \item Angles (in their own coordinate system) and Delays
 %   \item 4*100MHz
%\end{itemize}
The \ac{bs} sends \ac{OFDM} downlink signals to the receiver with a period of $\Delta$ seconds, which can arrive at the receiver directly, termed as the \ac{los} path, or reflected or diffused by the surfaces then reach the receiver, termed as \ac{nlos} paths, or via both \ac{los} and \ac{nlos} paths. When considering the transmissions of $N_{\text{OFDM}}$ \ac{OFDM} symbols with $S$ subcarriers, we can express the received signal for the $g$-th \ac{OFDM} symbol at time $k$ and the $\kappa$-th subcarrier as \cite{heath2016overview} 
\begin{align}
    {y}_{\kappa,g,k}&= \boldsymbol{w}_{\text{UE},g,k}^{\mathsf{H}} \underset{\boldsymbol{H}_{\kappa,k}} {\underbrace{\sum _{i=1}^{I_{k}}\rho_{k}^{i}\boldsymbol{a}_{\text{R}}(\boldsymbol{\theta}_{k}^{i})\boldsymbol{a}_{\text{T}}^{\mathsf{H}}(\boldsymbol{\phi}_{k}^{i})e^{-\jmath 2\pi \kappa \Delta_f \tau_{k}^{i}}}}  \boldsymbol{f}_{\text{BS},g,k} p_{\kappa,g} +  \boldsymbol{w}_{\text{UE},g,k}^{\mathsf{H}} \boldsymbol{n}_{\kappa,g,k},\label{eq:FreqObservation} 
\end{align}
where $p_{\kappa,g}$ denotes the pilot signal, $\boldsymbol{f}_{\text{BS},g,k}$ denotes the precoder at the transmitter,  $\boldsymbol{w}_{\text{UE},g,k}$ denotes the combiner at the receiver,
$\boldsymbol{H}_{\kappa,k}$ denotes the channel matrix, $\boldsymbol{a}_{\text{T}}(\cdot)$ denotes the steering vector at the transmitter side, $\boldsymbol{a}_{\text{R}}(\cdot)$ denotes the steering vector at the receiver side, $\Delta_f$ denotes the subcarrier spacing, and $\boldsymbol{n}_{\kappa,g,k}$ denotes white Gaussian noise across the receiver antenna arrays. In the propagation channel, there are $I_{k}$ paths in total, including both \ac{los} ($i=1$) and \ac{nlos} ($i>1$) paths, and each path can be described by a complex channel gain $\rho_{k}^{i}$, a \ac{TOA} $\tau_{k}^{i}$, an \ac{AOA} pair $\boldsymbol{\theta}_{k}^{i}=[\theta_{\text{az},k}^{i},\theta_{\text{el},k}^{i}]^{\mathsf{T}}$ in azimuth and elevation, and an \ac{AOD} pair $\boldsymbol{\phi}_{k}^{i}=[\phi_{\text{az},k}^{i},\phi_{\text{el},k}^{i}]^{\mathsf{T}}$ in azimuth and elevation. These channel parameters can be determined by the geometric relationships among the \ac{ue}, \ac{bs} and the \ac{IP} of the path if exists. It is important to note that $I_{k}$ is usually different from the number of visible landmarks, as a surface could generate more than one path.

\subsection{Measurement Model}
Given the received downlink signals ${y}_{\kappa,g,k}$, the channel estimator (introduced later in Section \ref{Section:channel_estimation}) can provide estimates of the channel parameters, i.e., \ac{TOA}, \ac{AOD}, \ac{AOA}, as measurements. However, all $I_{k}$ paths usually cannot have their corresponding measurements, due to limited resolution and imperfections of the channel estimator.
If we assume the $i$-th path has its corresponding measurement $\boldsymbol{z}_k^{i}$ and the measurement noise is a zero-mean Gaussian, $\boldsymbol{z}_k^{i}$  follows 
\begin{equation}\label{measurment_model}
    \boldsymbol{z}_k^{i} = 
    \boldsymbol{h}(\boldsymbol{s}_{k},\boldsymbol{s}_{\text{BS}},\boldsymbol{p}_{\text{IP},k}^{i} ) + \boldsymbol{r}_{k}^{i},
\end{equation} 
where $\boldsymbol{p}_{\text{IP},k}^{i}$ is its \ac{IP}, $\boldsymbol{r}_{k}^{i}\sim \mathcal{N}\left(\boldsymbol{0},\boldsymbol{R}_{k}^{i}\right)$ is the measurement noise, and $\boldsymbol{h}(\boldsymbol{s}_{k},\boldsymbol{s}_{\text{BS}},\boldsymbol{p}_{\text{IP},k}^{i} )= [{\tau}_{k}^{i},({\boldsymbol{\theta}}_{k}^{i})^{\mathsf{T}},({\boldsymbol{\phi}}^{i}_{k})^{\mathsf{T}}]^{\mathsf{T}}$ is the measurement function, representing the geometric relations among the \ac{ue}, \ac{bs} and the \ac{IP}, which is defined in Appendix \ref{GeometricRelationships}. Thus, we have \begin{align}
f(\boldsymbol{z}_{k}^{i}|\boldsymbol{p}_{\text{IP},k}^{i},\boldsymbol{s}_{k})=\mathcal{N}(\boldsymbol{z}_{k}^{i};\boldsymbol{h}(\boldsymbol{s}_{k},\boldsymbol{s}_{\text{BS}},\boldsymbol{p}_{\text{IP},k}^{i} ),\mathbf{R}_k^i) \label{likelihood}.
\end{align}
In principle, $\boldsymbol{p}_{\text{IP},k}^{i}$ does not exist, if it is a \ac{los} path, i.e., $i=1$. However, we still use $\boldsymbol{h}(\boldsymbol{s}_{k},\boldsymbol{s}_{\text{BS}},\boldsymbol{p}_{\text{IP},k}^{i} )$ for both \ac{los} and \ac{nlos} cases, to make the representation consistent.

\section{System Components}
\label{sec:Components}

The deployed 5G mmWave positioning system consists of a transmitter antenna array system at the \ac{bs} side and a receiver antenna array system at the \ac{ue} side operated at 27.2 GHz. 
The \ac{bs} broadcasts known reference signals to the \ac{ue}, which performs channel estimation, positioning, and mapping calculations.
Additionally, the \ac{ue} is equipped with a high-accuracy \ac{gnss} receiver, which provides the position and orientation of the \ac{ue} at any given time to be used as ground truth.
In this section, details about the system components are described.

\subsection{BS Description}
%\begin{itemize}
%    \item 1-1.5 column, 2-3 figures
%    \item BS Description
%    \item Antenna array
%    \item Deployment uncertainties in position and orientation
%    \item transmit signal
%    \item beam pattern (let Ericsson decide, we don't have, only can have public, like URA structure )
%    \item PRS
%\end{itemize}
% Placeholder text from FUSION paper. To be revised! [Peter]

%\PH{I note that we have not mentioned the carrier frequency. Should we do that in the BS section or will we tabulate any parameters including Fc?}

The \ac{bs} consists of a tower-mounted Antenna Integrated Radio unit for mmWave 5G NR (Ericsson AIR 5322), referred to as an \ac{AAS}, connected to a baseband unit located in a remote building. 
The baseband unit is responsible for controlling the reference signal generation, scheduling and mapping the signal to the radio resources, as well as controlling which beam to be used for the transmission at the \ac{AAS}.  
Central to the operation of the positioning system are the spatial properties of the beams at the output of the \ac{AAS}, as well as the properties of the transmitted reference signals, which are described below. 

\begin{figure}
    \centering
    \includegraphics[width=0.6\linewidth]{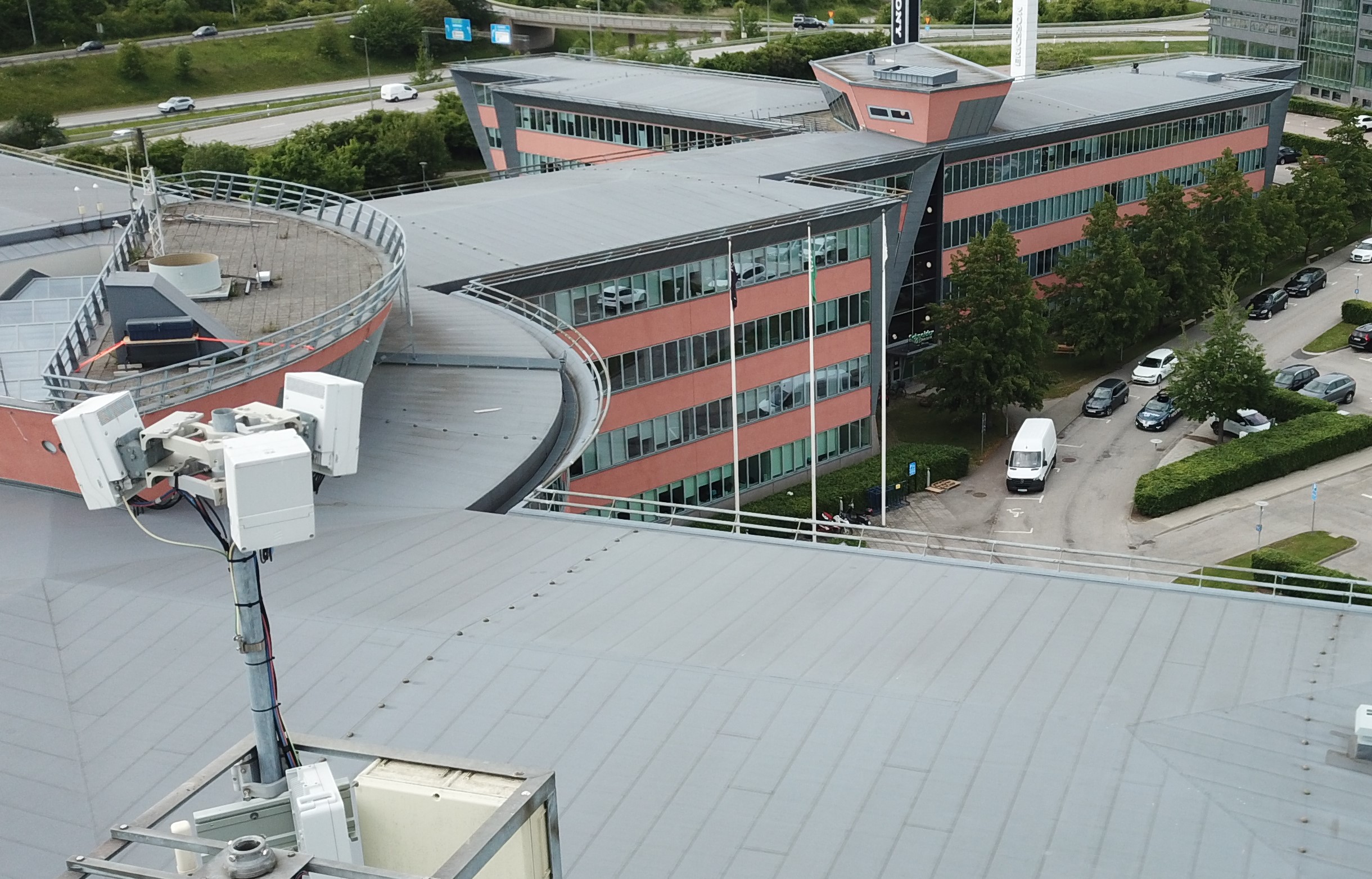}
    \caption{The tower \ac{bs} site, with three Ericsson AIR 5322 Antenna Integrated
Radio unit mounted to cover the surroundings. The unit facing the parking area was the one used during the experiments.}
    \label{fig:mounted_platform}
\end{figure}

\subsubsection{Antenna System Description}
\begin{figure}
    \centering
    \includegraphics[width=0.7\linewidth]{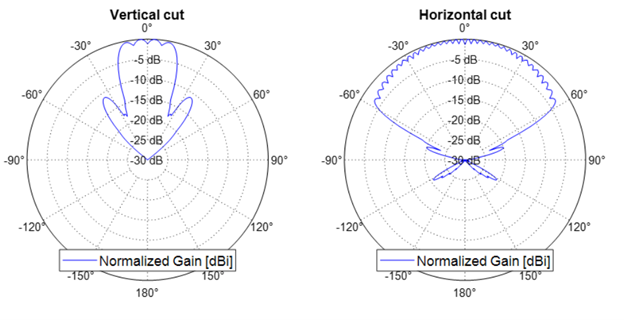}
    \caption{Horizontal and vertical cuts of the \ac{bs} beam pattern envelope.}
    \label{fig:BSbeams}
    \vspace{-4mm}
\end{figure}

%TODO: VERIFY IF WE WERE RUNNING WITH SPLIT ARRAY, IE 8x24 ANTENNAS
The \ac{AAS} is equipped with a phased array antenna module with a total of $16\times24$ dual-polarized antenna elements. 
Analog beamforming is implemented, creating a set of $4\times34$ (elevation $\times$ azimuth) available beams. 
The coverage area of the beam set is approximate $\pm~60^{\circ}$ in azimuth and $\pm~15^{\circ}$ in elevation, as illustrated in Fig.~\ref{fig:BSbeams}.
The characteristics of the individual beams show some variation depending on direction. However, the 3 dB beamwidth of the center beam is approximately $4.1^{\circ}$ in azimuth and $10.4^{\circ}$ in elevation.
%, and the aggregated beamwidth is approximate $\pm~60^{\circ}$ in azimuth and $\pm~15^{\circ}$ in elevation, as illustrated in Fig.~\ref{fig:BSbeams}.%and in the configuration used in the demo the \ac{AAS} can produce $4\times34$ beams. 
 In the custom-built beam-controlling software, all 136 beams are activated sequentially, with each beam being repeated over multiple \ac{OFDM} symbols, sweeping all beams with a periodicity of $40~\text{ms}$. 

For communication purposes, detailed information on the individual beam patterns is generally not required. 
Instead, information allowing for cell planning for data communication is available.
That is, $\boldsymbol{a}_{\text{T}}^{\mathsf{T}}(\boldsymbol{\phi}_{k}^{i}) \boldsymbol{f}_{\text{BS},g,k}$ is only known approximately (see Fig.~\ref{fig:BSbeams}), which may preclude the use of super-resolution methods for channel estimation. 
For the work presented in this paper, any detailed characterization of the produced beams has not been performed. Instead, we rely on the beam properties designed and obtained through simulations which, in general, show good correspondence. 
%However, it may not be the case for high accuracy positioning exploiting
%angular information of the incoming radio wave, which is derived using beam directions and underlying beam
%shape. In particular, for communication $\boldsymbol{a}_{\text{T}}^{\mathsf{T}}(\boldsymbol{\phi}_{k}^{i}) \boldsymbol{w}_{\text{BS},g,k}$ is only known approximately (see Fig.~\ref{fig:BSbeams}), while neither $\boldsymbol{a}_{\text{T}}(\boldsymbol{\phi}_{k}^{i})$ or $\boldsymbol{w}_{\text{BS},g,k}$ are characterized precisely, precluding the use of super-resolution methods. Hence, any characterization of the deployed \ac{AAS} and the produced beams have not been performed, instead we rely on the beam properties designed and obtained through simulations which in general show a good correspondence.   

\subsubsection{Reference Signal Description}\label{Sec_Sys_Ref}
To allow channel estimation at the \ac{ue}, the \ac{CSI-RS} is transmitted. 
The \ac{CSI-RS} is implemented according to the 3GPP NR 5G standard \cite{ts:38211-3gpp20}, configured to produce a pseudorandom sequence and mapped to QPSK symbols which are transmitted on every fourth subcarrier. The \ac{CSI-RS} is transmitted on each of the four phase-locked component carriers, producing a total bandwidth of $4\times100$ MHz. With 120 kHz subcarrier spacing and the standardized guard band between carriers, a total of $4\times198$ subcarriers are available for channel estimation.   
Additionally, to allow the \ac{ue} to perform time and frequency synchronization towards the \ac{bs}, \ac{SSB} \cite{ts:38211-3gpp20} is transmitted on dedicated beams at the beginning of the beam sweep. The \ac{SSB} contains synchronization signals and system information, which enable a \ac{ue} to perform synchronization and perform random access procedures for network connection.

\subsubsection{Deployment} \label{Sec_deploy}
The \ac{AAS} used in the experiment is part of a three-sector site, as seen in Fig.~\ref{fig:mounted_platform}, targeting omni-directional communication coverage. The two AASs not part of the experiment are configured to operate in a different frequency band. The site is situated at the top of a radio tower approximately $21~\mathrm{m}$  above the ground, with the \ac{AAS} mounted in a northbound direction overlooking a rooftop and a larger parking area surrounded by buildings. The AAS is down-tilted by $12^{\circ}$ to form the coverage area. Along with the coordinates of the base of the tower, it is possible to form the BS state $\boldsymbol{s}_\text{BS}$ vector. 
Even though the state values could be surveyed to arbitrary accuracy in theory, errors are inevitable in practice. 
% Even though the state values could, in theory, be surveyed to arbitrary accuracy, errors are, in practice, inevitable. 
Using sites that are intended to be used for communication, rather than accurate positioning, only coarse state information may be available. 
This is true for the deployment used in this experiment, where explicit surveying of the site has not been performed, and instead, over-the-air parameter estimation or calibration is performed as discussed in Section \ref{sec:bs_cal}.

\subsection{UE Description}
%\begin{itemize}
%    \item 2-3 columns, 2-3 figures
%    \item deployment of the array on the vehicle and corresponding uncertainties in position and orientation
%    \item Any other information regarding the vehicle you think is relevant
%    \item describe the hardware, antenna arrays
%    \item Describe the receiver beams
%    \item Describe uncertainties related to the array and beams and synchronization
%    \item the signal structure
%    \item the signal acquisition
%    \item signal processing steps
%    \item relevant uncertainties (eg. synchronization)
%    \item Different synchronization, like the synchronization between GPS and receiver, between BS and receiver (interpolation, compensate the bias with accurate clock, Meeting with Simon)
%\end{itemize}

\begin{figure}
    \centering
    \includegraphics[trim={40cm 25cm 30cm 15cm},clip,angle=-90,width=0.5\linewidth]{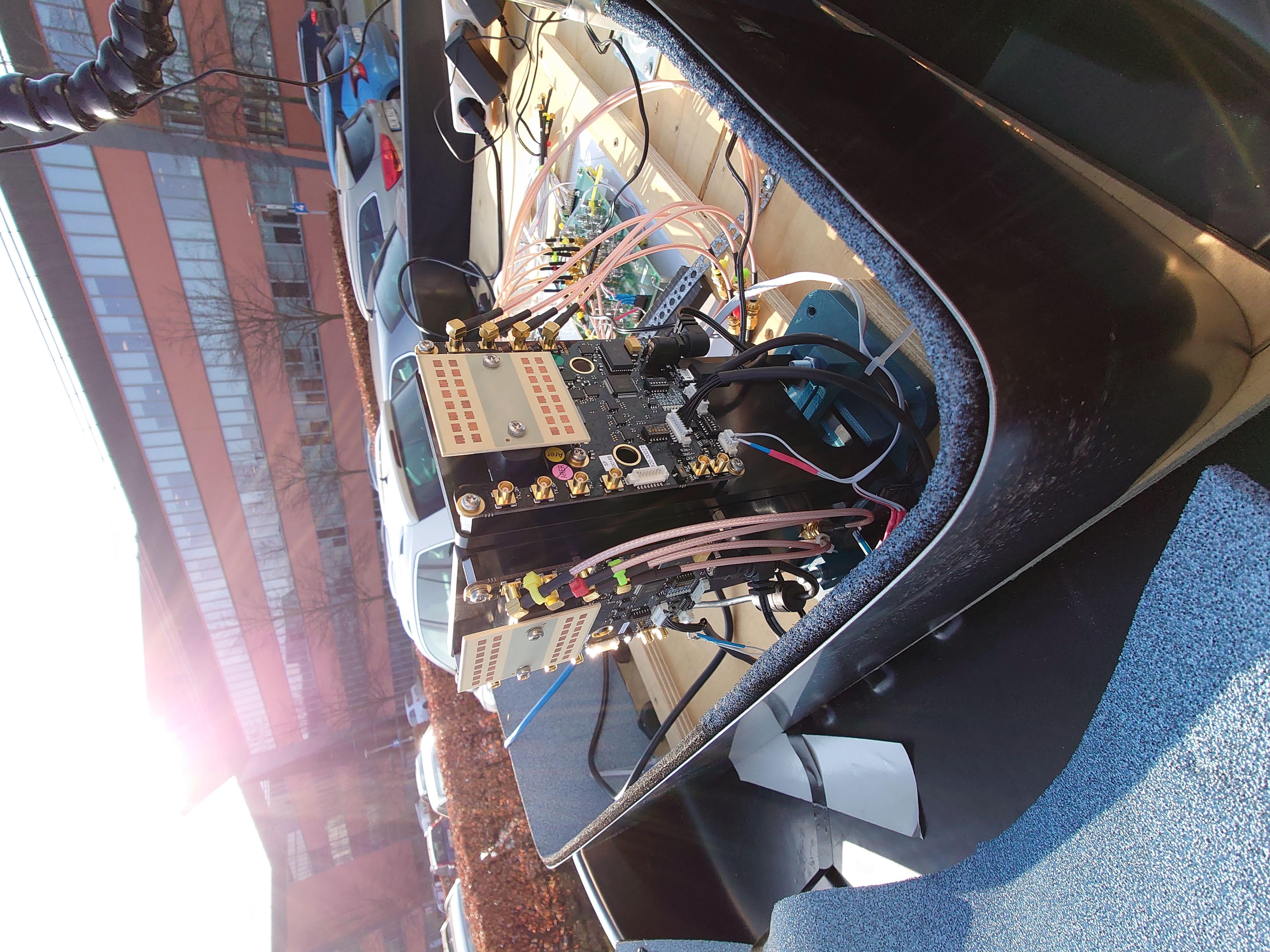}
    \caption{The two radio modules, with the \acf{RFSoC} in the background, mounted on the top of the car.}
    \label{fig:mounted_platform_ue}
\end{figure}

The \ac{ue} is based on the Sivers Semiconductors STP02800 platform. To increase the covered angle, the platform has been modified to run with two receiver channels. For the measurements, the platform was mounted on the roof of a test vehicle, see Fig~\ref{fig:mounted_platform_ue}. The radios were mounted perpendicular to each other with one facing forward and the other to the left of the car. Therefore, when the \ac{ue} drives towards the \ac{bs} from north to south, or alongside the \ac{bs} azimuth coverage from east to west, it could be assured that one of the radios lies within the \ac{los} while the other one could increase the chance of capturing the reflections.
Each radio is equipped with 2 URA antenna modules of size $2 \times 8$, where only one of them serves at the time, and the other is used as a backup. %With the beamsteering capability in the azimuth angle, There are 21 predefined beams covering the range $\pm45^{\circ}$.        

\begin{figure}
    \centering
    \includegraphics[width=0.6\linewidth]{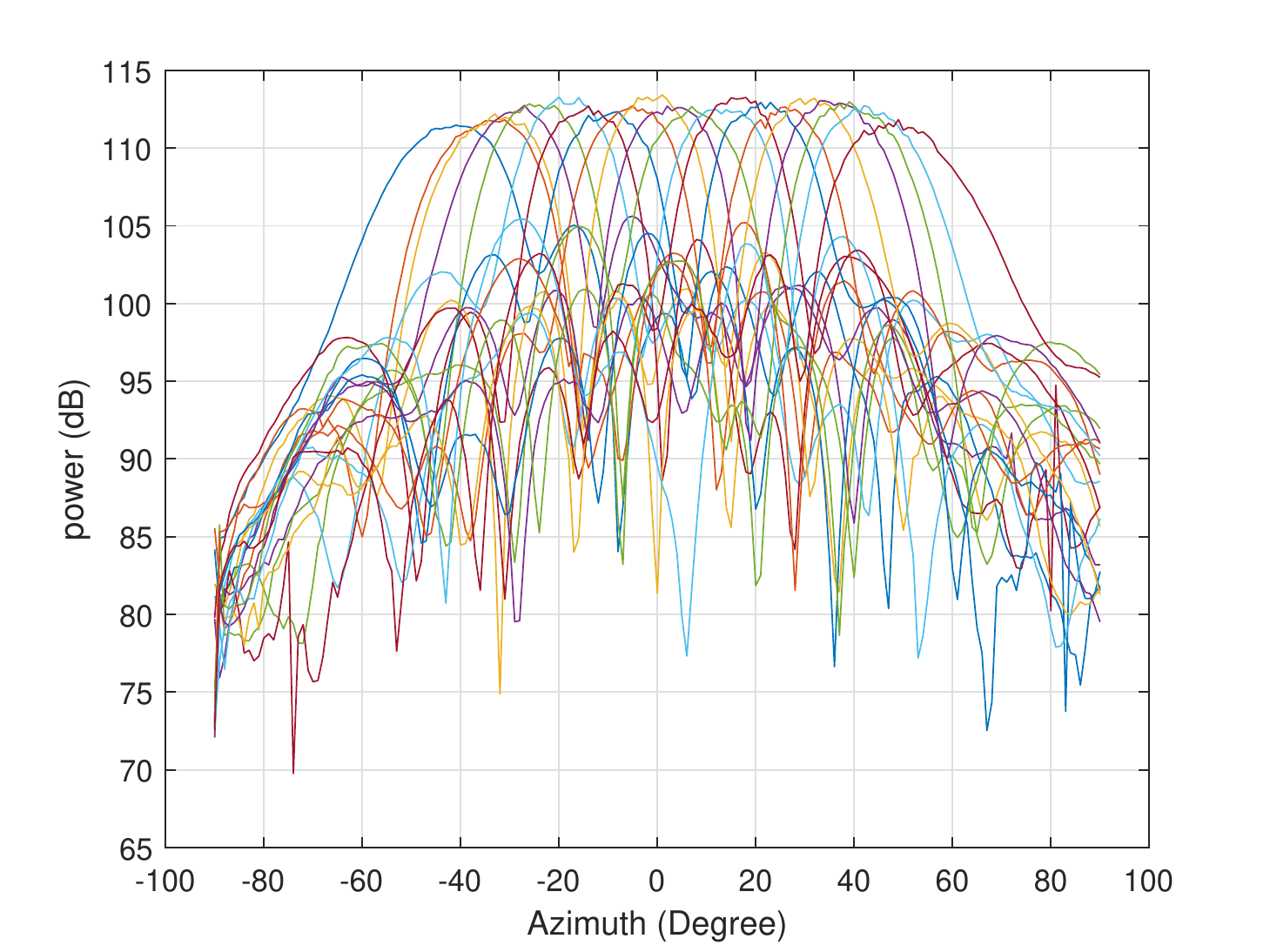}
    \caption{The UE beam patterns are characterized by the antenna responses measurement in a chamber at 27.2 GHz and predefined antenna weighting factors.}
    \label{fig:UE_Beams}
\end{figure}

 Unlike the BS side, the antenna response of the UE radio receiver has been characterized in a chamber with absorbers. It was done by $1^{\circ}$ of resolution in azimuth at 27.2 GHz with 400 MHz bandwidth. The Beam patterns were subsequently generated by applying the predefined antenna weighting vectors as shown in Fig~\ref{fig:UE_Beams}. Considering the overlapping of some UE beams, it might not be necessary to sweep all of them during the measurement.

As described earlier, the transmitted signal from the \ac{bs} consists of two parts. The first part consists of \ac{SSB} data for synchronization, and the second part consists of \ac{CSI-RS} data used for positioning. The first step in the calibration is to estimate the frequency offset between the \ac{bs} and \ac{ue}, and do the symbol timing synchronization. These steps are performed simultaneously using the \ac{PSS} part of the \ac{SSB}. The final step is the symbol synchronization which is achieved with the \ac{SSS} and \ac{PBCH-DMRS}. Once the synchronization is done, the \ac{RFSoC} keeps the synchronization for the rest of the measurements, meaning that the synchronization only has to be done once before starting the measurements. To be able to keep the synchronization, the clock on the \ac{RFSoC} has to be synchronized with the clock on the \ac{bs}. This was achieved by the use of an external rubidium clock providing the \ac{RFSoC} with a \ac{PPS} signal. With this solution, the \ac{RFSoC} was able to keep the synchronization with the \ac{bs} within a few samples.%$\pm2$~samples.

Once the synchronization is done, the data collection can begin. For each of the \ac{bs} beams, data was collected for $15$ of the $21$ available \ac{ue} beams. The used beam indices were $1, 3, 4, 5, 7, 8, 9, 11, 13, 14, 15, 17, 18, 19~\text{and}~21$. Each beam combination was measured for a duration of $2$ \ac{OFDM} symbols, which corresponds to around $18~\mathrm{\mu s}$. Each measurement lasted for $40$~ms and the data collected was downloaded from the \ac{RFSoC} and stored on a computer. Once the data is stored, a new measurement will start, resulting in a measurement taking place approximately every 3rd second. For each measurement, the current \ac{GPS} position was also saved to be used as a reference position.

\subsection{Ground-truth System}
%The test campaign used the Oxford RT3003G v2 system, referred to as OxTS, for the collection of ground truth trajectory. The system combines \ac{IMU}, comprised by three accelerometer and three gyroscopes. The OxTS can achieve $1~\text{cm}$ accuracy positioning by utilizing the \ac{RTK}. The \ac{RTK}  corrected \ac{GNSS} to provide a very accurate ground truth of the position of the ego vehicle. The Oxford RT3003G v2 system for this purpose. The system provides up to 1 cm accuracy positioning by combining the \ac{RTK} corrections with the \ac{GNSS} with the onboard \ac{IMU} and gyroscopes. The OXTS system provides outputs for latitude and longitude position, accelerations in three dimensions as well as rotations in three dimensions, with an update frequency of 100 Hz.

The test campaign used the Oxford RT3003G v2 system \cite{oxts}, referred to as OXTS, for the collection of ground-truth trajectory, which can achieve very accurate ground truth of the position of the ego vehicle down to $0.1~\text{m}$. The system consists of an \ac{IMU}, comprised of three accelerometers and three gyroscopes, a \ac{gnss} receiver, and a real-time on-board processor. A Kalman filter is used to fuse the IMU and GNSS inputs, which provides a real-time estimate of the position. The system also utilizes \ac{RTK} corrections provided via the SWEPOS network-\ac{RTK} service \cite{swepos} to further enhance the position accuracy, which is a technique used for resolving common errors in the GPS signal created by, for example, the atmosphere. By deploying a static \ac{bs} at a known position, an analysis of the carrier wave can be achieved and corrections can be broadcasted to the mobile rover in the area, hence making very accurate positioning achievable. Table \ref{tab:oxts_uncertainty} summarizes the average measurement uncertainties reported by the OXTS during the test.

The OXTS was installed in the vehicle's trunk, centered over the rear wheel axis. The system used a dual antenna setup to enhance the heading accuracy of the measurements during static measurements. The primary antenna was installed on top of the trunk of the vehicle. The secondary antenna was mounted on the engine hood. As the measurements from the OXTS were intended to describe the position and the orientation of the UE rather than the vehicle itself, all measurements were translated from the position and the orientation of the OXTS to the position and orientation of the UE antenna. The translation was based on manual measurements of the distances and rotations between the OXTS system to the UE on the roof of the vehicle. %, to the primary antenna and then to the UE on the roof of the vehicle. Measurements were done in (x,y,z) where x is the forward direction of the vehicle, y is to the left-hand side of the vehicle and z is the upwards direction.
% OxTS - What does it measure, what is RTK, how it works and what benifits it brings
% Vehicle installation, where is the OxTS mounted and how does it relate to the position of the UE (pos and rot)
% How we measured the distance between the UE and OxTS
% Figures... [I have some alternatives, perhaps exchange/complement visualization by Fan's pictures]

\begin{table}[h]
    \centering
    \caption{Average measurement uncertainties reported by OXTS during the test.}
    \label{tab:oxts_uncertainty}
    \begin{tabular}{|c|c|c|c|c|c|}
    \hline
     $\alpha_{\text{UE},k}$& $\beta_{\text{UE},k}$ &
    $\gamma_{\text{UE},k}$ &$x_{\text{UE},k}$ & $y_{\text{UE},k}$ & $z_{\text{UE},k}$\\ \hline
     $0.060^\circ$ & $0.052^\circ$ & $1.136^\circ$ & $0.194~\mathrm{m}$ &  $0.187~\mathrm{m}$ 
     &  $0.245~\mathrm{m}$\\
     \hline
    \end{tabular}
\end{table}

%\subsection{Beamspace Measurements}

\section{Channel Parameter Estimation, Positioning and Mapping} \label{sec:localization_mapping}

This section discusses how to obtain the channel parameters and how to perform positioning and mapping. The \ac{bs} calibration is first introduced, then the positioning algorithm using only \ac{los} is described, followed by  the positioning and mapping algorithm using both \ac{los} and \ac{nlos} paths. % for synchronized and asynchronized cases.

\subsection{Channel Parameter Estimation}\label{Section:channel_estimation}

At the CSI-RS transmission stage, both the BS and UE apply the beam-sweeping mechanism. Considering the number of beams on each side and the signal numerology, it was possible to send $30$ OFDM  symbols per each BS beam, which make sure all the $15$ UE beams are captured within $2$ ODFM symbols per UE beam. Taking the average over the $2$ received OFDM symbols per beam pair results in $G = 15\times 136$ OFDM symbols for each measurement. Therefore, we denote $\boldsymbol{\vartheta}_{g,k}$ and $\boldsymbol{\varphi}_{g,k}$ as the local beamforming angles for the $g$-th OFDM symbol at the UE and the BS sides, corresponding to $\boldsymbol{w}_{\text{UE},g,k}$ and $\boldsymbol{f}_{\text{BS},g,k}$, respectively. Based on the collected beamspace measurements $y_{\kappa, g, k}$ and the pilot signal $p_{\kappa,g}$ as described in Section~\ref{Sec_Sys_Ref}, the beamspace channel over the $\kappa$-th subcarrier, i.e., $h_{\kappa, g, k} = \boldsymbol{w}_{\text{UE},g,k}^\top \boldsymbol{H}_{\kappa, k}\boldsymbol{f}_{\text{BS}, g, k}$ is given by
\begin{equation}
    \hat h_{\kappa,g,k} = \frac{p_{\kappa,g}^*\hat y_{\kappa, g, k}}{|p_{\kappa,g}|^2} = h_{\kappa,g,k} + \frac{p_{\kappa,g}^*\hat n_{\kappa, g, k}}{|p_{\kappa,g}|^2}.
    \label{eq:beamspace_channel}
\end{equation}

Without completely calibrated complex beam pattern responses, a simple beam sweeping channel estimator can be designed to find the strongest beam pairs across the total $G$ transmissions. With the estimated beamspace channel in~\eqref{eq:beamspace_channel}, the strongest propagation path at the time $k$ can be obtained as
\begin{equation}
    \hat g_k = \argmax_g \sum_\kappa|\hat h_{\kappa, g,k}|^2,
\end{equation}
and the associated AOA and AOD can be obtained as  $\hat{\boldsymbol{\theta}}_k = \boldsymbol{\vartheta}_{\hat g, k}$ and $\hat{\boldsymbol{\phi}}_k = \boldsymbol{\varphi}_{\hat g, k}$. Considering the UE board has limited resolution on elevation, we describe the AOA estimation as $\hat \theta_k = [\theta_{\text{az}, k},0]^{\mathsf{T}}$ with setting the covariance of the AOA elevation very large later. Further refinement can be done by interpolation of adjacent beams. To do this, we first formulate a weighting tensor $\bar{\mathcal{\boldsymbol{H}}}\in \mathbb{R}^{15\times 4\times 34}$ at time $k$ with each element as
\begin{equation}
\bar{h}_{g, k} = \bar{h}_{g_1, g_2, g_3, k} = [\bar{\mathcal{\boldsymbol{H}}}_{k}]_{g_1, g_2, g_3} =  \sum_\kappa|\hat h_{\kappa, g,k}|^2,
\label{eq:tensor_channel}
\end{equation}
where the OFDM symbol index $g$ has a one-to-one mapping to indices $g_1, g_2, g_3$ of $\bar{\mathcal{\boldsymbol{H}}}_{k}$, which correspond to the indices of the UE beam, the BS elevation beam, and the BS azimuth beam at the $g$-th transmission, respectively.
% of the tensor space $\bar{\mathcal{\boldsymbol{H}}}_{k}$, and $g_1, g_2, g_3$ 
The refined angle estimation, taking AOD for example, can be calculated as 
\begin{equation}
    \hat{\boldsymbol{\phi}}_k = \frac{\sum_{g_2 \in \mathcal{G}_2}\sum_{g_3 \in \mathcal{G}_3} \bar{h}_{\hat g_1, g_2, g_3, k}\boldsymbol{\varphi}_{\hat g_1, g_2, g_3, k}}
    {\sum_{g_2 \in \mathcal{G}_2}\sum_{g_3 \in \mathcal{G}_3} \bar{h}_{\hat g_1, g_2, g_3, k}},
\end{equation}
where $\hat g_1$ is the index of the UE beam corresponding to the strongest propagation path $\hat g_k$, $\mathcal{G}_2$ and $\mathcal{G}_3$ contain the adjacent beams at the second and third dimension of the tensor, e.g., $G_3 = \{\hat g_3-1, \hat g_3, \hat g_3+1\}$ if $1<\hat g_3<15$.

For each beam-pair received signal, the phase changes linearly with the subcarrier index as indicated in \eqref{eq:FreqObservation}. However, the true phases are wrapped within $[0, 2\pi)$, and the delay for the strongest beam pair $\hat{g}$ can be estimated as
\begin{equation}
    \hat{\tau}_{k} = \argmin_{\tau} |\sum_{\kappa} {\hat{h}_{\kappa,\hat{g}_k,k}} e^{j 2 \pi \kappa \Delta_f \tau}|^2.
    \label{eq:delay_estimation}
\end{equation}
% \begin{equation}  
% \hat{\tau}_{k}^{i} = -\frac{\hat a_{k}}{2\pi \Delta_f},
% \end{equation}
% where
% \begin{equation}  
%         [\hat{a}_{k}, \hat{m}_{k}] = \argmin_{(a_{k},m_{k})} \sum_{\kappa}\| \angle{\hat{h}_{\kappa,\hat{g},k}} - a_{k}\kappa - m_{k}\|^2.
% \end{equation}
The AOAs/AODs and delays of other paths can be obtained in a similar way, providing the estimated channel parameters $\hat{\boldsymbol{z}}_k^i = [\hat \tau_k^i, (\hat{\boldsymbol{\theta}}_k^i)^{\mathsf{T}}, (\hat{\boldsymbol{\phi}}_k^i)^{\mathsf{T}}]^{\mathsf{T}}$.

In case only downlink measurements are used (see Section \ref{sec:TDOApositioning}), the UE clock bias is an unknown that should be estimated along with the UE position (see Appendix \ref{GeometricRelationships}). If the downlink measurement corresponds to the return part of an RTT protocol, the clock bias can be removed and the estimated TOA can be directly related to the distance between BS and UE. In this case, errors are due to the combination of clock stability, signal quality, and multipath resolution. The RTT model will be used in Section \ref{los_localization} and Section \ref{sec:RTT-NLOS}.

\subsection{BS Calibration}
\label{sec:bs_cal}
%\begin{itemize}
%    \item 6D BS calibration
%    \item MAP(ML) estimator
%    \item Data choose
%\end{itemize}

%In principle, the ground truth of the \ac{bs} state $\boldsymbol{s}_\text{BS}$ is given, and the \ac{ue} can directly use $\boldsymbol{s}_\text{BS}$ and measurements to localize itself. However, it is expected that the provided $\boldsymbol{s}_\text{BS}$ is not accurate enough and contains biases, as it is usually manually measured or the deployed position and orientation may change slightly after a long time exposure outside. This can cause significant errors, especially at a large distance. 
As discussed in Section~\ref{Sec_deploy}, explicit surveying of the site has not been performed. The \ac{bs} state $\boldsymbol{s}_\text{BS}$ is not provided with sufficient accuracy for positioning purposes.
Therefore, we need to calibrate the \ac{bs} state $\boldsymbol{s}_\text{BS}$ before doing positioning or other further processes. To do this, we utilize \ac{los} paths at different \ac{ue} locations, where the \ac{ue} positions and orientations at those locations are provided by the OXTS system. The \ac{nlos} paths are not used even though they contain the information on $\boldsymbol{s}_{\text{BS}}$, as the corresponding \ac{IP} of each \ac{nlos} path is unavailable. Moreover, we do not use the delay information of \ac{los} paths either, as the delay is affected by the clock bias. Although the BS and the UE are synchronized by the synchronization process, the clock bias between the BS and the UE is still time-varying and may suffer from non-negligible variations over time.

When the \ac{ue} arrives at a new location at time step $k$, the OXTS system provides its position $\boldsymbol{p}_{\text{UE},k}$, and the orientation $\boldsymbol{\psi}_{\text{UE},k}$. The channel estimator provides a group of measurements, and the measurement
associated with the \ac{los} path is denoted by $\boldsymbol{z}_{k}^{1}$, which is determined by selecting the strongest and shortest path as follows: ${\boldsymbol{z}}_{k}^{1} =  \{\boldsymbol{z}_{k}^{i}: \min_i \{\tau_k^i\} \}$. We pick up the AOA and the AOD of the \ac{los} path, denoted as $\check{\boldsymbol{z}}_{k}^{1}=[\boldsymbol{z}_{k}^{1}]_{2:5}$. 
% We only use the angle information of $\boldsymbol{z}_{k}^{1}$, which is denoted as $\check{\boldsymbol{z}}_{k}^{1}$. 
When considering the $K$ different locations, the posterior of $\boldsymbol{s}_{\text{BS}}$ given all $K$ groups of $\boldsymbol{p}_{\text{UE},k}$, $\boldsymbol{\psi}_{\text{UE},k}$ and $\check{\boldsymbol{z}}_{k}^{1}$ can be derived as 
\begin{align}
f(\boldsymbol{s}_\text{BS} |&\boldsymbol{p}_{\text{UE},1:K},\boldsymbol{\psi}_{\text{UE},1:K},\check{\boldsymbol{z}}_{1:K}^{1}) \propto \label{posterior_bs}f(\boldsymbol{s}_\text{BS})\prod_{k=1}^K f(\boldsymbol{p}_{\text{UE},k}) f(\boldsymbol{\psi}_{\text{UE},k})f(\check{\boldsymbol{z}}_{k}^{1}|\boldsymbol{p}_{\text{UE},k},\boldsymbol{\psi}_{\text{UE},k},\boldsymbol{s}_\text{BS}),
\end{align}
where $f(\boldsymbol{s}_{\text{BS}})$ is a uniform prior for $\boldsymbol{s}_\text{BS}$ within a small area centered at the provided ground truth, and $f(\boldsymbol{p}_{\text{UE},k})$ and $f(\boldsymbol{\psi}_{\text{UE},k})$ denote the densities of the $\boldsymbol{p}_{\text{UE},k}$ and $\boldsymbol{\psi}_{\text{UE},k}$, respectively, with the mean measured by the OXTS system,  and covariances as shown in Table \ref{tab:oxts_uncertainty}, and $f(\check{\boldsymbol{z}}_{k}^{1}|\boldsymbol{s}_{\text{UE},k},\boldsymbol{\psi}_{\text{UE},k},\boldsymbol{s}_\text{BS})$ is the likelihood of $\check{\boldsymbol{z}}_{k}^{1}$, which is equivalent to taking the angular part of \eqref{likelihood}. Then, $\boldsymbol{s}_\text{BS}$ can be estimated by maximize  \eqref{posterior_bs}
\begin{align}
 \hat{\boldsymbol{s}}_\text{BS} &=\argmax_{\boldsymbol{s}_\text{BS}}f(\boldsymbol{s}_\text{BS} |\boldsymbol{p}_{\text{UE},1:K},\boldsymbol{\psi}_{\text{UE},1:K},\check{\boldsymbol{z}}_{1:K}^{1}).\label{max_pos}
\end{align}
Considering the uniform and Gaussian distributions in \eqref{posterior_bs}, it is equivalent to write \eqref{max_pos} as 

\begin{align}
 \hat{\boldsymbol{s}}_{\text{BS}} =\argmin_{\boldsymbol{s}_{\text{BS}}}\sum_{k=1}^K &\left( \check{\boldsymbol{h}}_{k}^{1} - \check{\boldsymbol{z}}_{k}^{1} \right)^\top (\check{\boldsymbol{R}}_{k}^{1})^{-1}  \left( \check{\boldsymbol{h}}_{k}^{1} - \check{\boldsymbol{z}}_{k}^{1} \right)  \label{LS}\\
\text{s.t.} \quad  & \boldsymbol{s}_\text{BS} \in 
\mathbb{S}_{\text{BS}}\nonumber,
\end{align}
where $\mathbb{S}_{\text{BS}}$ denotes a small space centered around the uncalibrated \ac{bs} state, and $\check{\boldsymbol{h}}_{k}^{1}$ and $\check{\boldsymbol{R}}_{k}^{1}$ are the angular parts of ${\boldsymbol{h}}_{k}^{1}$ and ${\boldsymbol{R}}_{k}^{1}$, respectively. Since \eqref{LS} does not have a closed-form solution, $\hat{\boldsymbol{s}}_\text{BS}$ is approximated iteratively. The problem is initialized using the uncalibrated state, and the gradient-based trust-region-reflective algorithm is used to solve the problem.

\subsection{Positioning based on LOS path only}\label{los_localization}

%\begin{itemize}
%        \item Describe the used information
%        \item ML estimator
%        \item LOS pick-up
%    \end{itemize}

After the \ac{bs} calibration, a more accurate \ac{bs} state can be acquired, and we directly  denote the calibrated \ac{bs} state as $\boldsymbol{s}_\text{BS}$ from now on. When knowing $\boldsymbol{s}_\text{BS}$, we can use measurements provided by the channel estimator to localize the \ac{ue}, which is to estimate $\boldsymbol{p}_{\text{UE},k}$ in this paper. The most straightforward way is to utilize the \ac{los} path. If $\boldsymbol{\psi}_{\text{UE},k}$ is unknown to the \ac{ue}, $\boldsymbol{\theta}_{k}^{1}$ cannot be used, as it also depends on the orientation of the \ac{ue}. In this case, only $\tau_{k}^{1}$ and $\boldsymbol{\phi}_{k}^{1}$ can be used, and $\boldsymbol{p}_{\text{UE},k}$ has a closed-form solution as 
\begin{equation}\label{positioning_syn}
    \boldsymbol{p}_{\text{UE},k} = \boldsymbol{p}_\text{BS} + d_{k}^{1}\underset{{\boldsymbol{u}^{1}_{\text{BS},k}}}{\underbrace{\boldsymbol{R}_{\text{BS}} ^{\textsf{T}}\begin{bmatrix} \cos\left({\phi}_{\text{az},k}^{1} \right)\cos\left({\phi}_{\text{el},k}^{1} \right) \\ 
    \sin\left({\phi}_{\text{az},k}^{1} \right)\cos\left({\phi}_{\text{el},k}^{1} \right) \\
\sin\left({\phi}_{\text{el},k}^{1}\right)\end{bmatrix}}},
\end{equation}
where $d_{k}^{1}=( \tau_{k}^{1} - b_{k} )c$ is the propagation distance of the \ac{los} path, and $\boldsymbol{R}_{\text{BS}}^{\textsf{T}}$ is the rotation matrix from local coordinate system of the \ac{bs} to the global reference. This approach works only if the clock bias is eliminated by the RTT protocol, %the synchronization problem between the transmitter and the receiver is solved 
and the clock bias is quite stable during the estimation procedure. %, so that the clock bias is given or can be estimated. 
However, if the clock bias varies seriously with time, \eqref{positioning_syn} will result in large positioning errors. Although $\boldsymbol{\theta}_{k}^{1}$ cannot be used in these cases, it is still possible to do single-\ac{bs} positioning, as long as we know at least one dimension of $\boldsymbol{p}_{\text{UE},k}$, e.g., the height of the \ac{ue} $z_{\text{UE},k}$. Knowing $z_{\text{UE},k}$ is a reasonable assumption, especially in the urban vehicular scenario, where the vehicle moves on the ground, and the height does not change much. Then, the 2D position of the \ac{ue} on the x-y domain can be computed by
%\begin{align}\label{positioning_unsyn}
%    &\begin{bmatrix} x_{\text{UE},k} \\ 
%    y_{\text{UE},k} \end{bmatrix}= \begin{bmatrix} x_{\text{BS}} \\ 
%    y_{\text{BS}} \end{bmatrix} + \\& \qquad \frac{ z_{\text{UE},k} - z_{\text{BS}} }{\sin\left(\boldsymbol{\theta}_{\text{el},k}^{1} \right)}[\boldsymbol{R}_{\text{BS}} ^{\textsf{T}}]_{1:2,1:3}\begin{bmatrix} \cos\left(\boldsymbol{\theta}_{\text{az},k}^{1} \right)\cos\left(\boldsymbol{\theta}_{\text{el},k}^{1} \right) \\ 
%    \sin\left(\boldsymbol{\theta}_{\text{az},k}^{1} \right)\cos\left(\boldsymbol{\theta}_{\text{el},k}^{1} \right) \\
%\sin\left(\boldsymbol{\theta}_{\text{el},k}^{1}\right)\end{bmatrix}.\nonumber
%\end{align}
\begin{align}\label{positioning_unsyn}
    &\begin{bmatrix} x_{\text{UE},k} \\ 
    y_{\text{UE},k} \end{bmatrix}= \begin{bmatrix} x_{\text{BS}} \\ 
    y_{\text{BS}} \end{bmatrix} +  \frac{ z_{\text{UE},k} - z_{\text{BS}} }{[\boldsymbol{u}^{1}_{\text{BS},k}]_{3}}[\boldsymbol{u}^{1}_{\text{BS},k}]_{1:2}.
\end{align}
In an ideal case, where the clock bias is stable and provided, and $\boldsymbol{\psi}_{\text{UE},k}$ is known, $\boldsymbol{p}_{\text{UE},k}$ can be acquired by
\begin{align}
 \hat{\boldsymbol{p}}_{\text{UE},k} =\argmin_{\boldsymbol{p}_{\text{UE},k}} \left( {\boldsymbol{h}}_{k}^{1} - {\boldsymbol{z}}_{k}^{1} \right)^\top ({\boldsymbol{R}}_{k}^{1})^{-1}  \left( {\boldsymbol{h}}_{k}^{1} - {\boldsymbol{z}}_{k}^{1} \right)  \label{LS_pos_ideal},
\end{align}
where $\boldsymbol{h}_{k}^{1}$ is the shorthand of $\boldsymbol{h}(\boldsymbol{s}_{k},\boldsymbol{s}_{\text{BS}})$. Note that $\boldsymbol{p}_{\text{UE}}$ is included in $\boldsymbol{s}_{\text{BS}}$ by definition.

\subsection{Positioning based on LOS and NLOS }

%    \begin{itemize}
%        \item Data Association
%        \item clustering algorithm
%        \item ML estimator for sensor station
%        \item How to map the environment
%    \end{itemize}
In Section \ref{los_localization}, only the \ac{los} path is used to position the \ac{ue} every single time snapshot, and all the \ac{nlos} paths are not utilized. These \ac{nlos} paths contain information on the  \ac{ue} position, which is beneficial to positioning. Since \ac{nlos} paths also depend on the surrounding environment, it is possible to both localize the \ac{ue} and map the environment by utilizing all \ac{los} and \ac{nlos} paths. To solve this positioning and mapping problem, the most straightforward way is to first solve the \ac{ML} problem as
\begin{align}
 \hat{\boldsymbol{p}}_{\text{UE},k},\hat{b}_{k},[\hat{\boldsymbol{p}}_{\text{IP},k}^{i}]_{i=2}^{\hat{I}_{k}}=\argmax_{\boldsymbol{p}_{\text{UE},k},b_{k},[\boldsymbol{p}_{\text{IP},k}^{i}]_{i=2}^{\hat{I}_{k}}}\prod_{i=1}^{\hat{I}_{k}}f(\boldsymbol{z}_{k}^{i}|\boldsymbol{s}_{k},\boldsymbol{s}_{\text{BS}},\boldsymbol{p}_{\text{IP},k}^{i})\label{ML_LM},
\end{align}
After getting all \acp{IP} $[\hat{\boldsymbol{p}}_{\text{IP},k}^{i}]_{i=2}^{\hat{I}_{k}}$, surfaces can be estimated from those points, as \acp{IP} of paths reflected/diffused by the same surface should be on the same surface.

\subsubsection{RTT-based Positioning} \label{sec:RTT-NLOS}
\begin{figure}
    \centering
    \includegraphics[width=0.6\linewidth]{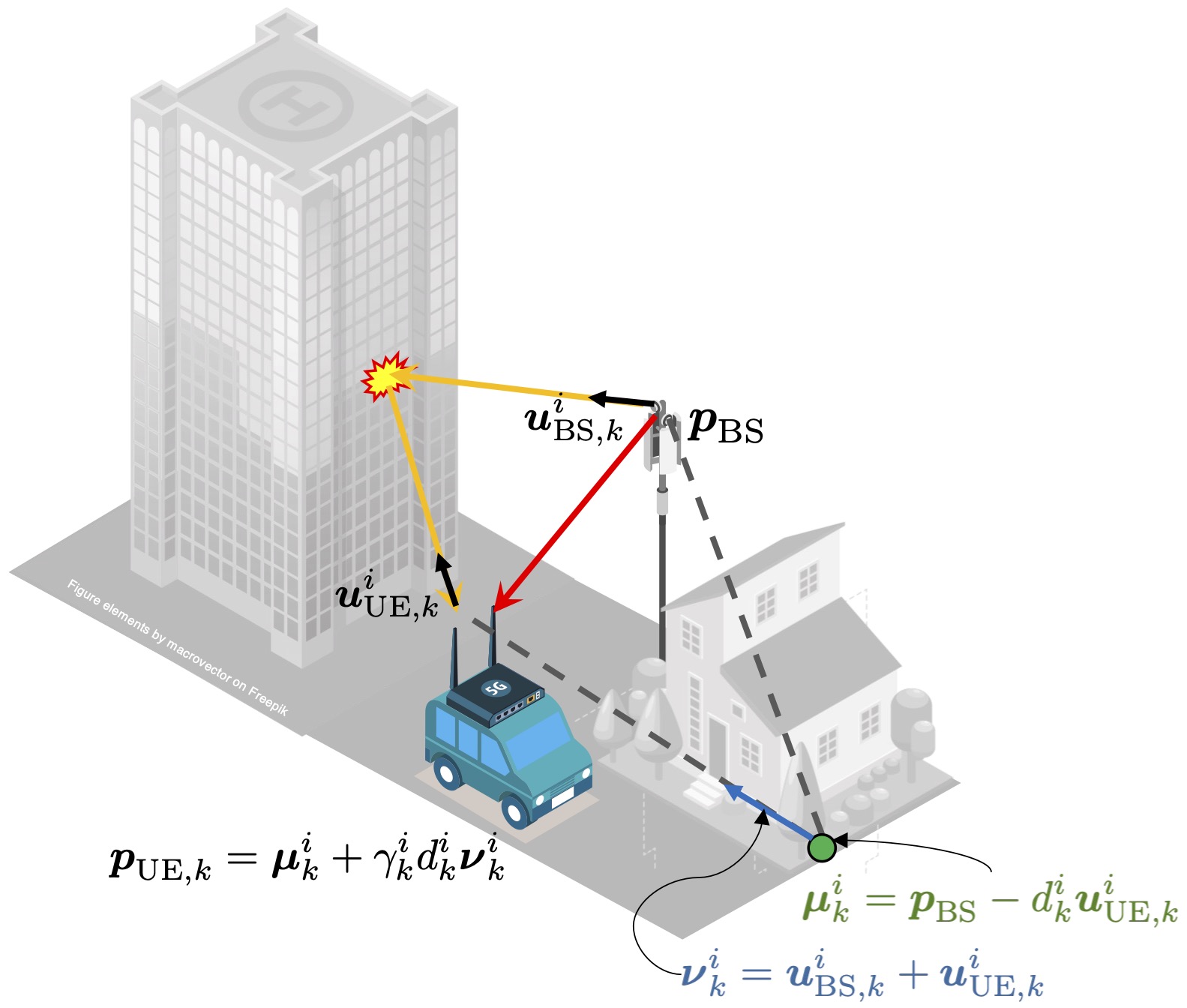}
    \caption{The visualization of \eqref{lines}, where $\boldsymbol{u}^{i}_{\text{BS},k}$, $\boldsymbol{u}^{i}_{\text{UE},k}$, $\boldsymbol{p}_\text{BS}$ and $ d_{k}^{i}$ are known,  $\gamma_{k}^{i}$ is unknown, and  $\boldsymbol{p}_{\text{UE},k}$ is on the line of $\boldsymbol{\mu}_{k}^{i}+\boldsymbol{\nu}_{k}^{i}$.}
    \label{fig:nlos_local}
\end{figure}

Even with an RTT-based protocol where $b_k$ can be assumed known, the high-dimensional optimization is required in \eqref{ML_LM}, which usually has high complexity. To avoid this, we propose a low-complex search-free positioning and mapping method that follows two steps. The first step is to localize the \ac{ue} using all paths. For a path $i$, we define a unit vector at the transmitter side $\boldsymbol{u}^{i}_{\text{BS},k}$ to represent the direction of the signal leaving the transmitter in the global coordinate system, denoted as
\begin{align}
    \boldsymbol{u}^{i}_{\text{BS},k} =\boldsymbol{R}_{\text{BS}} ^{\textsf{T}}\begin{bmatrix} \cos\left({\phi}_{\text{az},k}^{i} \right)\cos\left({\phi}_{\text{el},k}^{i} \right) \\ 
    \sin\left({\phi}_{\text{az},k}^{i} \right)\cos\left({\phi}_{\text{el},k}^{i} \right) \\
\sin\left({\phi}_{\text{el},k}^{i}\right)\end{bmatrix}.
\end{align}
Similarly, we define a unit vector at the receiver side $\boldsymbol{u}^{i}_{\text{UE},k}$ to represent the direction of the signal arriving at the receiver in the global coordinate system, denoted as
\begin{align}
\boldsymbol{u}^{i}_{\text{UE},k}=\boldsymbol{R}_{\text{UE},k} ^{\textsf{T}}\begin{bmatrix} \cos\left({\theta}_{\text{az},k}^{i} \right)\cos\left({\theta}_{\text{el},k}^{i} \right) \\ 
    \sin\left({\theta}_{\text{az},k}^{i} \right)\cos\left({\theta}_{\text{el},k}^{i} \right) \\
\sin\left({\theta}_{\text{el},k}^{i}\right)\end{bmatrix}.\label{ue_position_fra}
\end{align}
where $\boldsymbol{R}_{\text{UE},k}^{\textsf{T}}$ denotes the rotation matrix from the local coordinate system of the UE to the global reference.
Then, the \ac{ue} position can be given by
\begin{align}
\boldsymbol{p}_{\text{UE},k} = \boldsymbol{p}_\text{BS} + d_{k}^{i}\gamma_{k}^{i}\boldsymbol{u}^{i}_{\text{BS},k} - d_{k}^{i}(1-\gamma_{k}^{i})\boldsymbol{u}^{i}_{\text{UE},k},
\end{align}
where $d_{k}^{i}=( \tau_{k}^{i} - b_{k} )c$ denotes the corresponding propagation distance of the path, and $\gamma_{k}^{i}\in[0,1]$ represents the fraction of the propagation distance on the direction of $\boldsymbol{u}^{i}_{\text{BS},k}$,  which is unknown yet. By rearranging \eqref{ue_position_fra} and gathering known and unknown parts separately, we have
\begin{align}\label{lines}
\boldsymbol{p}_{\text{UE},k} = \underset{\boldsymbol{\mu}_{k}^{i}} {\underbrace{\boldsymbol{p}_\text{BS}-d_{k}^{i}\boldsymbol{u}^{i}_{\text{UE},k}}} + \gamma_{k}^{i} d_{k}^{i}\underset{\boldsymbol{\nu}_{k}^{i}} {\underbrace{(\boldsymbol{u}^{i}_{\text{BS},k} +\boldsymbol{u}^{i}_{\text{UE},k})}},
\end{align}
which means $\boldsymbol{p}_{\text{UE},k}$ is $\gamma_{k}^{i}d_{k}^{i}|\boldsymbol{\nu}_{k}^{i}|$ away from the point $\boldsymbol{\mu}_{k}^{i}$ alongside the direction of $\boldsymbol{\nu}_{k}^{i}$, i.e., the direction of $\boldsymbol{u}^{i}_{\text{BS},k} +\boldsymbol{u}^{i}_{\text{UE},k}$, and it is obvious that $\boldsymbol{p}_{\text{UE},k}$ is on the determined line, as described in Fig.~\ref{fig:nlos_local}. There are $\hat{I}_{k}$ estimated paths, and each will determine a line. Therefore, $\boldsymbol{p}_{\text{UE},k}$
can be determined by finding the intersection points of these lines. This can be done by minimizing the cost function
\begin{align} \label{distance_to_lines}
 \hat{\boldsymbol{p}}_{\text{UE},k}=\argmin_{\boldsymbol{p}_{\text{UE},k}}\sum_{i=1}^{\hat{I}_{k}}\eta_{k}^{i}||\boldsymbol{p}_{\text{UE},k}-\boldsymbol{\mu}_{k}^{i}-(\bar{\boldsymbol{\nu}}_{k}^{i})^{\textsf{T}}(\boldsymbol{p}_{\text{UE},k}-\boldsymbol{\mu}_{k}^{i})\bar{\boldsymbol{\nu}}_{k}^{i}||^{2},
\end{align}
with $\eta_{k}^{i}$ denoting the strength of the $i$-th path, and $\bar{\boldsymbol{\nu}}_{k}^{i}=\boldsymbol{\nu}_{k}^{i}/|\boldsymbol{\nu}_{k}^{i}|$ denoting the corresponding norm vector of the direction. The interpretation of \eqref{distance_to_lines} is finding a $\boldsymbol{p}_{\text{UE},k}$ that results in the smallest sum of the distance between $\boldsymbol{p}_{\text{UE},k}$ and the line \eqref{lines} of each path. There is a closed-form solution for \eqref{distance_to_lines}, which is the least-square solution given by
\begin{align} \label{pos_least_square}
 \hat{\boldsymbol{p}}_{\text{UE},k}=\Big(\sum_{i=1}^{\hat{I}_{k}}\eta_{k}^{i}(\boldsymbol{I}-\bar{\boldsymbol{\nu}}_{k}^{i}(\bar{\boldsymbol{\nu}}_{k}^{i})^{\textsf{T}})\Big)^{-1}\sum_{i=1}^{\hat{I}_{k}}\eta_{k}^{i}(\boldsymbol{I}-\bar{\boldsymbol{\nu}}_{k}^{i}(\bar{\boldsymbol{\nu}}_{k}^{i})^{\textsf{T}})\boldsymbol{\mu}_{k}^{i}.
\end{align}

\subsubsection{TDOA-based Positioning} \label{sec:TDOApositioning}
The solution of \eqref{pos_least_square} requires a known $b_{k}$. It is still possible to simultaneously localize the \ac{ue} and synchronize it with the \ac{bs}. To do it, we firstly reconstruct \eqref{lines} as
\begin{align}\label{lines_new}
\boldsymbol{p}_{\text{UE},k}-b_{k}c\boldsymbol{u}^{i}_{\text{UE},k} = \underset{\tilde{\boldsymbol{\mu}}_{k}^{i}} {\underbrace{\boldsymbol{p}_\text{BS}-\tau_{k}^{i}c\boldsymbol{u}^{i}_{\text{UE},k}}} + \gamma_{k}^{i} d_{k}^{i}\underset{\boldsymbol{\nu}_{k}^{i}} {\underbrace{(\boldsymbol{u}^{i}_{\text{BS},k} +\boldsymbol{u}^{i}_{\text{UE},k})}}.
\end{align}
As we want to estimate both $\boldsymbol{p}_{\text{UE},k}$ and $b_{k}$, we can reformulate the optimization problem in \eqref{distance_to_lines} as
\begin{align} \label{distance_to_lines_new}
 &\hat{\boldsymbol{p}}_{\text{UE},k},\hat{b}_{k}=\argmin_{\boldsymbol{p}_{\text{UE},k},b_{k}}\sum_{i=1}^{\hat{I}_{k}}\eta_{k}^{i}||\boldsymbol{p}_{\text{UE},k}-b_{k}c\boldsymbol{u}^{i}_{\text{UE},k}-\tilde{\boldsymbol{\mu}}_{k}^{i}-(\bar{\boldsymbol{\nu}}_{k}^{i})^{\textsf{T}}(\boldsymbol{p}_{\text{UE},k}-b_{k}c\boldsymbol{u}^{i}_{\text{UE},k}-\tilde{\boldsymbol{\mu}}_{k}^{i})\bar{\boldsymbol{\nu}}_{k}^{i}||^{2}.
\end{align}
The closed-form solution of \eqref{distance_to_lines_new} is given by
\begin{align} \label{pos_bias_least_square}
\begin{bmatrix} \hat{\boldsymbol{p}}_{\text{UE},k} \\ \hat{b}_{k}\end{bmatrix}&=\Big(\sum_{i=1}^{\hat{I}_{k}}\eta_{k}^{i}(\boldsymbol{J}_{k}^{i})^{\textsf{T}}(\boldsymbol{I}-\bar{\boldsymbol{\nu}}_{k}^{i}(\bar{\boldsymbol{\nu}}_{k}^{i})^{\textsf{T}})\boldsymbol{J}_{k}^{i}\Big)^{-1}\sum_{i=1}^{\hat{I}_{k}}\eta_{k}^{i}(\boldsymbol{J}_{k}^{i})^{\textsf{T}}(\boldsymbol{I}-\bar{\boldsymbol{\nu}}_{k}^{i}(\bar{\boldsymbol{\nu}}_{k}^{i})^{\textsf{T}})\tilde{\boldsymbol{\mu}}_{k}^{i},
\end{align}
where $\boldsymbol{J}_{k}^{i}=[\boldsymbol{I},-c\boldsymbol{u}^{i}_{\text{UE},k}]$.

\subsection{Estimating \ac{IP} Locations (mapping)}
After getting $\hat{\boldsymbol{p}}_{\text{UE},k}$ and $\hat{b}_{k}$ (if required), we can compute each $\boldsymbol{p}_{\text{IP},k}^{i}$ for all \ac{nlos} paths ($i>1$) by
\begin{align}
 \hat{\boldsymbol{p}}_{\text{IP},k}^{i} =\argmin_{\boldsymbol{p}_{\text{IP},k}^{i}}\left( {\boldsymbol{h}}_{k}^{i} - {\boldsymbol{z}}_{k}^{i} \right)^\top ({\boldsymbol{R}}_{k}^{i})^{-1}  \left( {\boldsymbol{h}}_{k}^{i} - {\boldsymbol{z}}_{k}^{i} \right)  \label{LS_inc_points},
\end{align}
where $\boldsymbol{h}_{k}^{i}$ is the shorthand of $\boldsymbol{h}(\boldsymbol{s}_{k},\boldsymbol{s}_{\text{BS}},\boldsymbol{p}_{\text{IP},k}^{i} )$, and $\hat{\boldsymbol{p}}_{\text{UE},k}$ in \eqref{pos_least_square} is plugged into $\boldsymbol{h}_{k}^{i}$.  The optimization problem in \eqref{LS_inc_points} initialized by the intersection point of two lines determined by $(\boldsymbol{p}_\text{BS} + d_{k}^{i}\gamma_{k}^{i}\boldsymbol{u}^{i}_{\text{BS},k})$ and $(\hat{\boldsymbol{p}}_\text{UE,k} + d_{k}^{i}(1-\gamma_{k}^{i})\boldsymbol{u}^{i}_{\text{UE},k})$, which is given by
\begin{align}
 &\hat{\boldsymbol{p}}_{\text{IP},k}^{i} =\left((\boldsymbol{I}-\boldsymbol{u}^{i}_{\text{BS},k}(\boldsymbol{u}^{i}_{\text{BS},k})^{\textsf{T}})+(\boldsymbol{I}-\boldsymbol{u}^{i}_{\text{UE},k}(\boldsymbol{u}^{i}_{\text{UE},k})^{\textsf{T}})\right)^{-1} \label{initialization}\left((\boldsymbol{I}-\boldsymbol{u}^{i}_{\text{BS},k}(\boldsymbol{u}^{i}_{\text{BS},k})^{\textsf{T}})\boldsymbol{p}_\text{BS} +(\boldsymbol{I}-\boldsymbol{u}^{i}_{\text{UE},k}(\boldsymbol{u}^{i}_{\text{UE},k})^{\textsf{T}})\hat{\boldsymbol{p}}_\text{UE,k} \right),
\end{align}
where $\tau_{k}^{i}$ is not directly used. As there is no closed-form solution, to use all the measurement components and consider the uncertainty of the measurement, the optimization problem can be approximated by using the sigma point least square method, where a list of sigma points is created, Levenberg-Marquardt algorithm \cite{marquardt1963algorithm} is used to approximate the optimization problem for each sigma point, and the estimated state can be acquired by averaging over all optimization outputs of sigma points.

\section{Results} \label{results}
In this section, we introduce the test environment and display the channel estimation,  \ac{bs} calibration, \ac{los} positioning, and \ac{los} and \ac{nlos} positioning and mapping results by using test measurements. We also analyze some potential improvements by comparing the results using real test measurements with the results using the combination of parts of the test measurements and the simulated measurements.

\subsection{Test Environment}
\begin{figure}
    \centering
    \includegraphics[width=0.6\linewidth]{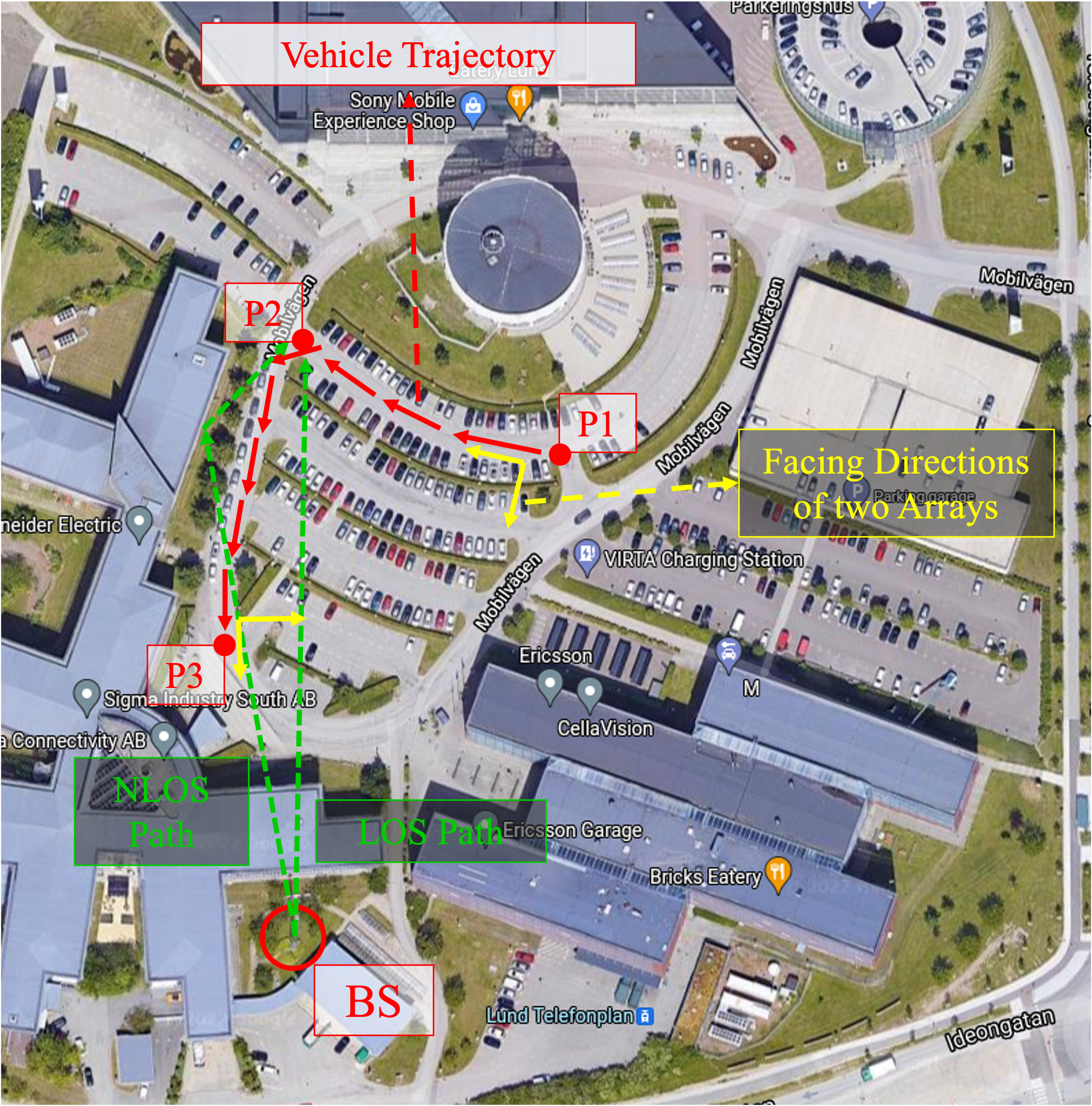}
    \caption{The test happened at Ideon scientific park, Lund. The \ac{bs} is mounted on a signal tower, visible at the red circle in the bottom left part of the figure. The vehicle slowly drove alongside the trajectory, shown as the red arrows in the figure, with the antenna arrays towards the \ac{bs}. The \ac{ue} received signals sent from the \ac{bs}. The maximum distance between the UE and the BS is about 130 m, and the minimum distance is about 85 m. The dashed green lines are examples of a \ac{los} and a \ac{nlos} at a specific \ac{ue} location. }
    \label{fig:google_map}
\end{figure}

%    \begin{itemize}
%        \item Parameters 
%        \item Location
%    \end{itemize}
The tests were carried out at Ideon scientific park, Lund, as shown in Fig.~\ref{fig:google_map}. The single \ac{bs} was fixed on the top of the signal tower at coordinate (55°42'58.6"N 13°13'32.9"E) with a height around $21~\mathrm{m}$, and the transmitter antenna array at the \ac{bs} was facing north with a $12^\circ$ tilt angle, see Fig.~\ref{fig:mounted_platform}.  There was no explicit surveying of the \ac{bs} performed before, and these measurements are some preliminary measurements that contain biases and need to be calibrated. The vehicle has its UE platform mounted on its roof with one of the two \acp{URA} facing to the \ac{bs}, as shown in Fig.~\ref{fig:mounted_platform_ue}. There was an absorption shield below the UE platform, which can avoid some ground reflections. The front \ac{URA} was aligned with the vehicle's direction. The relative positions and the orientations of the \ac{ue} \acp{URA} with respect to the \ac{ue}'s GPS point on the rear axis, in longitudinal, lateral, height, roll, pitch, and yaw, were measured manually onsite before starting the actual test. Although these measurements may contain some minor errors, those errors are negligible; thus, we ignored them in this paper. 

During the test, we first drove the vehicle randomly to many different locations in the parking area and kept it static to collect measurements for the \ac{bs} calibration purpose. After that, we did the test for positioning and mapping purposes. The vehicle was slowly driven alongside the trajectory described as the red arrows in Fig.~\ref{fig:google_map} for measurement collection. All these test points and the trajectory were in an effective measurement area, where the \ac{bs} can always be seen by the \ac{ue}, and there are large multi-story buildings that can have some reflections or diffusion.  The \ac{ue} continuously received downlink signals sent from the transmitter every a few seconds, which can arrive at the receiver both via the \ac{los} and possible \ac{nlos} paths. In the meantime, the GPS measured the ground truth of the \ac{ue} position and orientation. We repeated the same test for positioning and mapping purposes a few times. Note that although two rounds of the test should follow the same designed trajectory in principle, we cannot make the vehicle exactly follow the same trajectory every time. Therefore, each test still had different measurements and ground truths. After data collection, the received downlink signals were further processed and used for the \ac{bs} calibration and positioning and mapping purposes. The GPS measurements were used for the \ac{bs} calibration as well as the evaluation of positioning performances of different algorithms.

\subsection{Results and Discussion}
\subsubsection{Channel estimation}
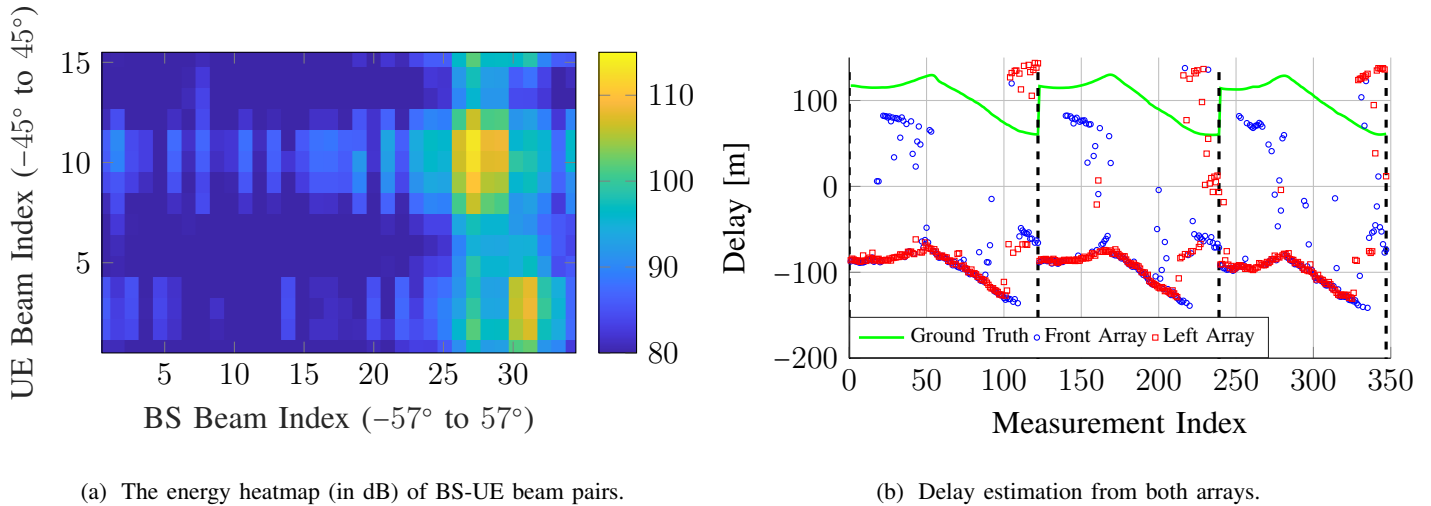
\begin{figure}[t!]
\subfigure[\,The energy heatmap (in dB) of BS-UE beam pairs.]{\input{Figures_without_ESPRIT/c_1}\label{fig:CE01}}
\subfigure[\,Delay estimation from both arrays.]{\input{Figures_without_ESPRIT/c_2_3times}\label{fig:CE02}}

% \subfigure[\,The delay of each beam.]{\input{5G mmWave Positioning/c_3}\label{beampair_delay}}
\caption{Channel estimation results for the vehicle trajectory in Fig.~\ref{fig:google_map}. (a) Beamspace channel matrix of a single measurement at location P2 with LOS and NLOS paths; (b) Delay estimations of the LOS path from two arrays (including clock bias, repeated 3 times).}
\label{fig:channel_estimation}
% \vspace{-5mm}
\end{figure}

We first visualize the beamspace channel of a specific position (i.e., P2) in terms of received signal strength by fixing BS elevation beams (i.e., $\bar{\boldsymbol{H}}_{[:, \hat g_2, :]}\in \mathbb{R}^{15\times 34}$), based on~\eqref{eq:beamspace_channel}--\eqref{eq:tensor_channel}, as shown in Fig.~\ref{fig:CE01}. The two strong beam pairs (27, 11) and (31, 3) can be clearly observed, which correspond to the LOS path and NLOS path, respectively. By processing all the measurements based on~\eqref{eq:delay_estimation} along the vehicle trajectory (i.e.,  solid red arrow from P1 to P3 in Fig.~\ref{fig:google_map}), the delay (with clock bias) estimations from both the front and left board for repeated three times of measurement are plotted in Fig.~\ref{fig:CE02}. {We can see that the clock bias does not drift (i.e., there is no increasing error over time), but still % . Please note that the drift means that the error increases over time. Even though clock bias does not drift, it is still time-varying and may 
suffers from non-negligible variations, due to  hardware imperfections, imperfect synchronization, and unresolved multipath. %Therefore, the clock bias can be compensated by the mean of the clock bias over time, and it performs well if the variation is not high.
} In the first part of the measurement, the left array has better visibility with fewer outliers (red square markers) than the front array. However, when the vehicle continues heading south (towards P3), the channel estimates from both boards are getting unstable as the transmitter and receiver are far away from the boresights of each other's antenna array. The results from Fig.~\ref{fig:CE02} show that the benefit from beamforming gain by using an antenna array (instead of an omnidirectional antenna) will cause coverage issues (e.g., both arrays cannot be seen by the BS when the vehicle is approaching P3), and a multi-panel mmWave transceiver array, as well as adaptive beamforming, could be adopted on the vehicle side to mitigate this effect.

\subsubsection{BS calibration}
After the implementation of the channel estimator on all received measurements, we first take all measurements for the \ac{bs} calibration purpose and apply the \ac{bs} calibration algorithm described in Section \ref{sec:bs_cal}. The calibrated \ac{bs} results are summarized in Table \ref{tab:bs_cal}, compared with the provided "ground truth". The calibrated 3D \ac{bs} position is $[0.78~\mathrm{m},0.73~\mathrm{m},18.66~\mathrm{m}]^{\textsf{T}}$ and calibrated Euler angle is $[1.25^\circ,9.92^\circ,73.45^\circ]^{\textsf{T}}$, while the uncalibrated position and orientation are $[0,0,21~\mathrm{m}]^{\textsf{T}}$ and  $[0,12^\circ,69^\circ]^{\textsf{T}}$, respectively. From the result, we observe that the provided \ac{bs} state gives a relatively good x-y position, but a relatively bad orientation and height, which is due to the orientation are relatively hard to measure when deployed and there was no explicit surveying of the site performed before. Moreover, we only know the radio tower is approximately 21 meters, but its global reference height is unknown. These biases can result in positioning errors.
\begin{table}[t]
    \centering
    \caption{The comparison between the calibrated \ac{bs} state and the uncalibrated \ac{bs} state.}
        \begin{tabular}{ |c|c|c| } 
         \hline
           State & Before calibration  & After calibration  \\ \hline 
         Position & $[0,0,21~\mathrm{m}]^{\textsf{T}}$ & $[0.78~\mathrm{m},0.73~\mathrm{m},18.66~\mathrm{m}]^{\textsf{T}}$
         \\ 
         Orientation  & $[0,12^\circ,69^\circ]^{\textsf{T}}$ & $[1.25^\circ,9.92^\circ,73.45^\circ]^{\textsf{T}}$  
         \\
         \hline
        \end{tabular} 
    \label{tab:bs_cal}
\end{table}

%We  implemented the \ac{bs} positioning algorithm on the channel estimates of the real measurements. We recall that since the UE and BS are not synchronized, the LOS path only provides information via the AOA and AOD. We find that the overall error is $24.82~\mathrm{m}$, based on the measurements from P1 and P2 in Fig.~\ref{fig:meaEnv}. This is because %that the delay cannot be used due to the serious clock drift, no resolution on \ac{AOA} elevation is provided, and  measurements are from two 2 different \ac{ue} locations. although there are 20 measurements in total, they are from only 2 different locations, hence measurements from the same locations are highly correlated. Only knowing the BS direction with respect to the two different UE locations, the \ac{bs} location is found by the intersection point of the two corresponding lines, which is sensitive to the angle errors, especially at a long distance. 

\begin{figure}
\centering
\input{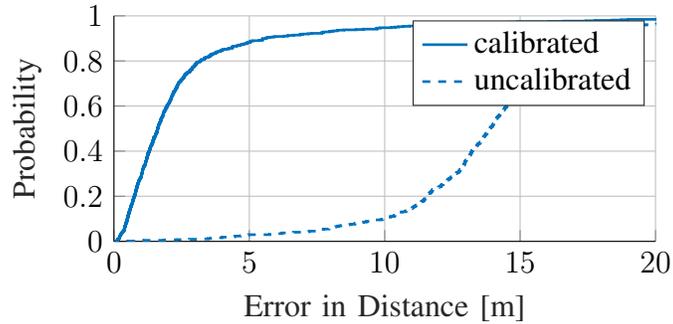}
\caption{\acp{CDF} of errors on the x-y plane of the LOS-only positioning algorithm using \ac{AOD} measurements assuming UE height is known, given the calibrated or uncalibrated \ac{bs} states.}
\label{fig:2D_cali}
% \vspace{-5mm}
\end{figure}

We now evaluate the benefits of the \ac{bs} calibration. We assume the height of the  \ac{ue} is known, and apply the positioning algorithm described as \eqref{positioning_unsyn} on the channel estimation results given the calibrated and uncalibrated \ac{bs} states, respectively, as shown in Fig.~\ref{fig:2D_cali}. Since the UE height is known, only the AOD is used. 
From the figure, we find that better positioning performance can be acquired when using the calibrated \ac{bs} state, as the solid line is always on the left side of the dashed line. This is because the positioning algorithms highly depend on the \ac{bs} state, and the biases of the \ac{bs} state lead to positioning errors, which are especially sensitive to the angle errors at a long distance and a small bias can cause a significant error. By using the calibrated \ac{bs} state, the positioning algorithm using only LOS can get below-5-meter accuracy with 88.3\% and  below-10-meter accuracy with 94.8\%, if the UE height is given.

\subsubsection{\ac{los} positioning}
\begin{figure}
\centering
\input{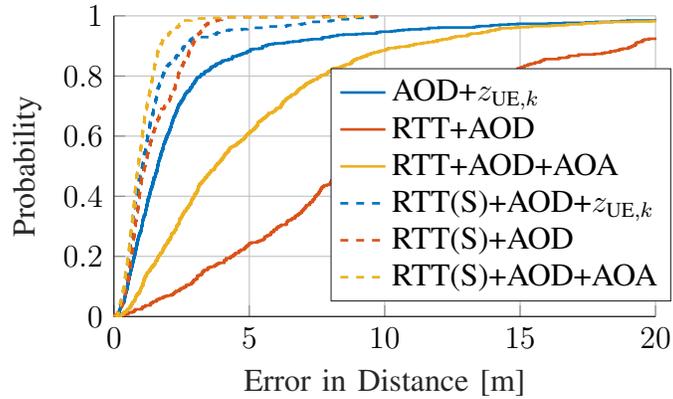}
\caption{\acp{CDF} of errors on the x-y plane of the three \ac{los} positioning algorithms different combinations of information, where "AOD+$z_{\text{UE},k}$" represents \eqref{positioning_unsyn}, "RTT+AOD" represents \eqref{positioning_syn}, "RTT+AOD+AOA" represents \eqref{LS_pos_ideal}. All methods use experimental data, except RTT(S), which uses synthetic RTT measurements.}
\label{fig:3D_los_without_esprit}
% \vspace{-5mm}
\end{figure}

After the \ac{bs} calibration, we then implement the \ac{los} positioning algorithm, described in Section \ref{los_localization}, on the channel estimation results of the measurements for the positioning purpose. To emulate RTT measurements, we remove a fixed clock bias from all the measurements and rely on the external rubidium clock to maintain a stable clock at the UE. 
{We only show the results on the x-y domain and the results on height estimates are omitted; as in a vehicular scenario, all vehicles move on the ground and the x-y position is more important than the height. However,  errors on the height follow similar tendencies as errors on the x-y domain.} We apply the positioning algorithm using experimental \ac{rtt} and \ac{AOD}, described as \eqref{positioning_syn} and the positioning algorithm using experimental \ac{rtt}, \ac{AOD} and \ac{AOA}, described as \eqref{LS_pos_ideal}, compared with the performance of the positioning algorithm using experimental \ac{AOD} and ground-truth UE height, and results are as solid lines shown in Fig.~\ref{fig:3D_los_without_esprit}. We observe that the performance of the algorithm using \ac{rtt} and \ac{AOD} is not as good as the performance of the algorithm using \ac{AOD} and ground-truth UE height, as below-10-meter accuracy is only 62.4\%, compared with 94.8\% using experimental \ac{AOD} and ground-truth UE height. 
{This degradation is due to the relatively poor RTT estimates, as seen in Fig.~\ref{fig:CE02}, which exhibit a \ac{MAE} of the estimated \ac{los} propagation distance of around 9.8 m. The error is attributed to a combination of clock instability, insufficient resolution of the beam sweeping channel estimation, as well so that inter-path interference due to limited resolution.} 
We can get better positioning performance than the algorithm using \ac{rtt} and \ac{AOD} when  also  \ac{AOA} is included, as the solid yellow line is always on the left side of the solid red line and the below-10-meter accuracy is 88.5\%. The reason is that more information is provided by the \ac{AOA}. However, it performs worse than the algorithm using  \ac{AOD} and ground-truth UE height, as the solid yellow line is on the right side of the solid blue line. The reason is that the unknown height adds one additional degree of freedom to the state, combined with the aforementioned RTT errors. To remove the impact of the delay measurements,  we  replace the experimental \ac{rtt} with synthetic measurements, which are generated by adding noise with $1~\mathrm{m}$ standard deviation to the ground truth propagation distances. The results of positioning algorithms using a mix of experimental and synthetic data are displayed as dashed lines in Fig.~\ref{fig:3D_los_without_esprit}, compared with the result when using pure experimental data. The more accurate RTT helps positioning algorithms to achieve better positioning performances, as the dashed lines are all on the left side of the solid lines, and the combination of synthetic RTT, AOD and AOA provides around 1.71 m positioning accuracy with 90\% of cases. Such performance levels can possibly be attained by high-resolution estimators. 
%Therefore, 
%
%\Yu{we can try to better solve the clock instability problem to get more accurate \ac{rtt}, as well as provide high-resolution channel estimators in the future}.
\begin{comment}
\begin{figure}
\centering
\input{Figures/error_los_more_mea.tex}
\caption{\acp{CDF} of errors on x-y plane of the three \ac{los} positioning algorithms.}
\label{fig:3D_los}
% \vspace{-5mm}
\end{figure}
\end{comment}

\subsubsection{\ac{los} and \ac{nlos} positioning and mapping}
\begin{figure}
    \centering
    \includegraphics[width=0.6\linewidth]{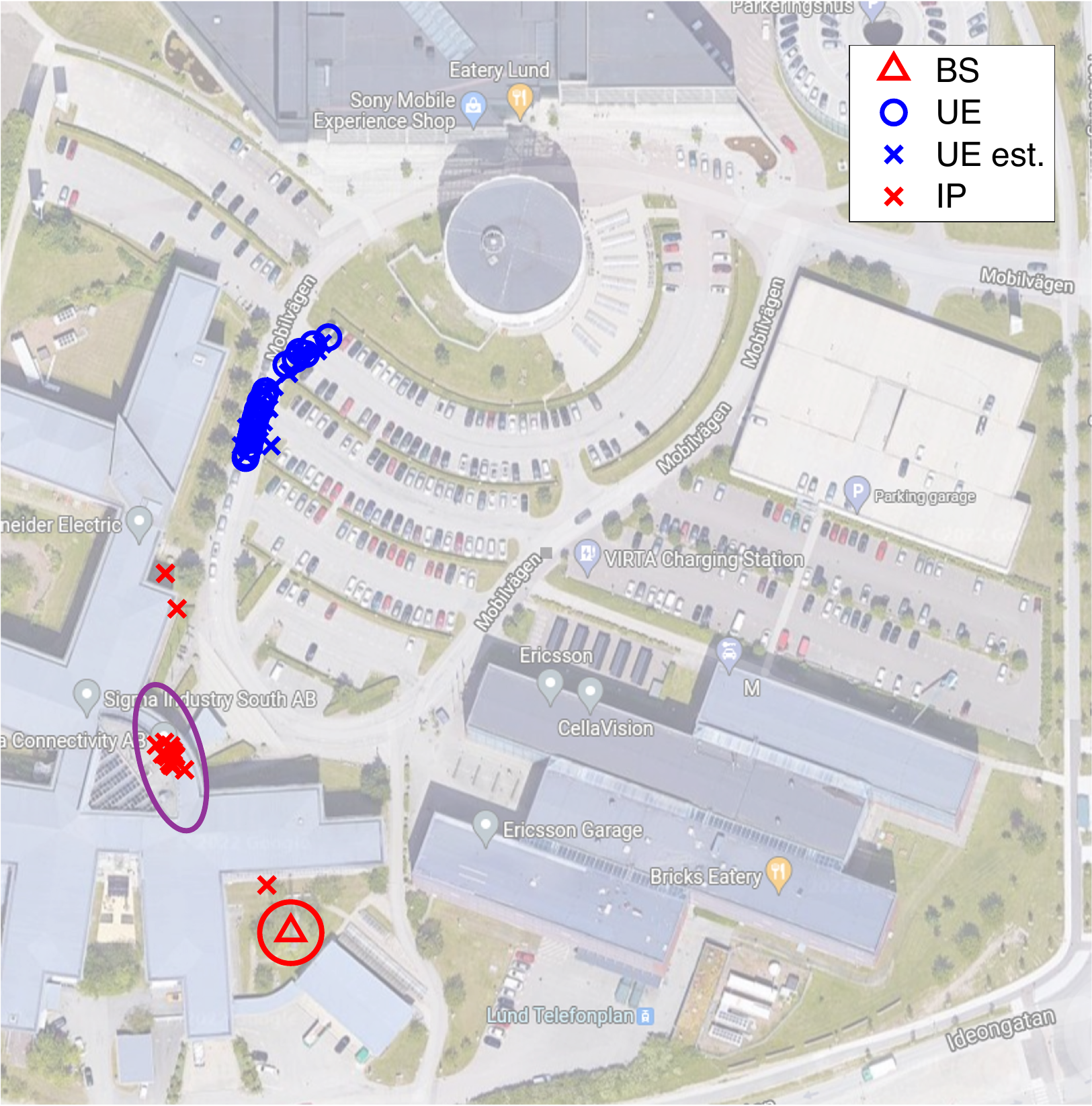}
    \caption{The visualization of the result of the \ac{los} and \ac{nlos} positioning and mapping algorithm evaluated at the selected 20 points where \ac{nlos} can be resolved if the clock bias is known.}
    \label{fig:nlos_lam}
\end{figure}

\begin{figure}
    \centering   \includegraphics[width=0.6\linewidth]{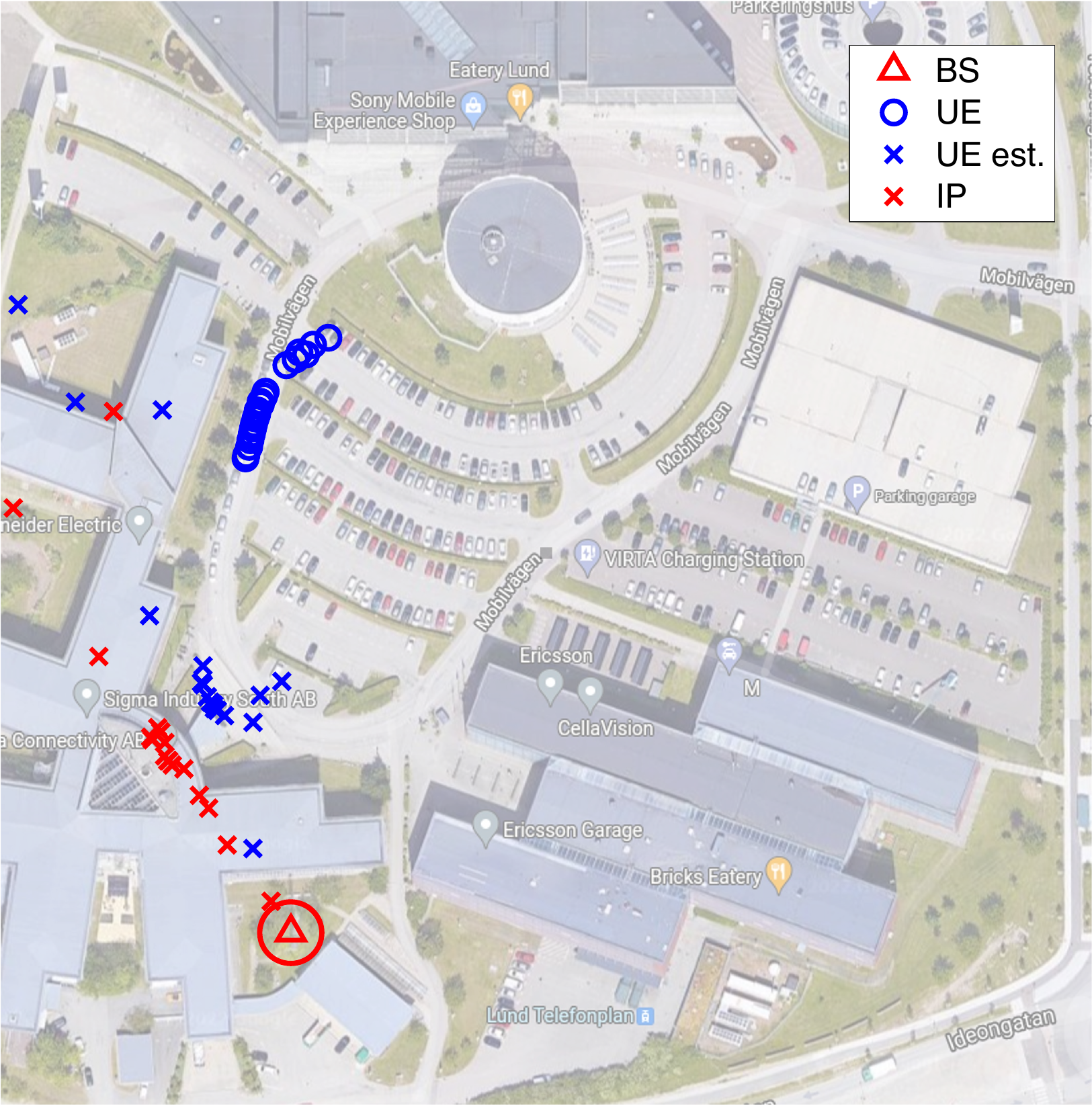}
    %\vspace{-3mm}
    \caption{The visualization of the result of the \ac{los} and \ac{nlos} positioning and mapping algorithm evaluated at the selected 20 points where \ac{nlos} can be resolved if the clock bias is unknown.}
\label{fig:nlos_lam_unsyn}
\end{figure}

In the previous sections, we have not considered \ac{nlos} paths, and we start to consider \ac{nlos} paths from now on. However, \ac{nlos} paths are not always visible at all \ac{ue} locations, as these paths are too weak to be resolved when the \ac{ue} are far away from landmarks. %As not all \ac{ue} locations have clear \ac{nlos} paths, 
We picked 20 points where some \ac{nlos} paths can be clearly observed. Then the \ac{los} and \ac{nlos} positioning algorithm (see \eqref{pos_least_square}) are implemented at these points and its performance is evaluated by the \ac{MAE} of the x-y plane, which is on average $3.98~\text{m}$ compared with $4.86~\text{m}$ when only \ac{los} paths are used for these points. The better positioning performance is acquired due to the help of the \ac{nlos} paths. After getting the estimated \ac{ue} positions, we then apply the mapping algorithm to get \acp{IP}. Fig.~\ref{fig:nlos_lam} summarizes the results of \ac{los} and \ac{nlos} positioning and mapping algorithm for the selected points. From Fig.~\ref{fig:nlos_lam}, we can see that the algorithm can localize the \ac{ue} even though there are some errors, as the blue crosses are close to blue circles. \acp{IP} of all \ac{nlos} are recovered by \eqref{initialization}, shown as red crosses in Fig.~\ref{fig:nlos_lam}. Although we do not have the ground truth of the \acp{IP}, 
the locations of these points in the ground-truth map are reasonable, which are on the surfaces of the buildings. From \acp{IP}, we can know where the surfaces are. Therefore, both positioning and mapping are achieved by utilizing the \ac{los} and \ac{nlos} paths. There are only a few \acp{IP} and only one surface can be mapped, shown as the magenta circle, which is because we do not have enough \ac{nlos} paths. However, if we could observe more \ac{nlos} paths, more landmarks can be mapped. However, this is for the synchronized case where the clock bias is known, we also implement the algorithms for the asynchronized case, as shown in Fig.~\ref{fig:nlos_lam_unsyn}. The positioning and mapping results are not good, because there is only one \ac{nlos} observed at each time point. Clock bias gets very large estimation error, as a few degrees errors in angles make large errors on bias estimation, which further leads to bad positioning and mapping permanence. %If the known the \acp{IP}, we can also implement \eqref{pos_bias_least_square_knowmap} to get the \ac{ue} position and clock bias. We then use the \acp{IP} in Fig.~\ref{fig:nlos_lam} to do this work, and the results are shown in Fig.~\ref{fig:nlos_lam_unsyn_knownmap}. We find noticeable errors in the estimations, which is due to the estimation of \acp{IP} are not good. Together with the \ac{los} path, only two paths can easily get very large errors on the joint state of \ac{ue} position and clock bias.

%\subsubsection{\ac{los} and \ac{nlos} mapping}
%\begin{figure}
%\centering
%\input{Figures/error}
%\caption{The CDF of errors for under different settings, and w.cal. represents with BS calibration, and w.o.cal represents without BS calibration.}
%\label{fig:cur_set}
% \vspace{-5mm}
%\end{figure}

\begin{comment}

\begin{figure}
    \centering   \includegraphics[width=\columnwidth]{Figures_without_ESPRIT/nlos_lam_noVA_unsyn_knownmap.pdf}
    \caption{The visualization of the result of the \ac{los} and \ac{nlos} positioning and mapping algorithm evaluated at the selected 20 points where \ac{nlos} can be resolved if the \acp{IP} are known and the clock bias is unknown.}
\label{fig:nlos_lam_unsyn_knownmap}
\end{figure}
\end{comment}

\section{Conclusions} \label{Conclusions}
5G mmWave signals are useful for \ac{ue} positioning for intelligent transport, due to the increased spatial and temporal resolution, compared to sub-6 GHz signals. Current solutions are based on using several \acp{bs} to determine the UE location, which poses significant demands in terms of 5G infrastructure. In this paper, we evaluated, through real-world demonstration, the ability to localize a \ac{ue} by a single \ac{bs}, operating according to the current release of the 3GPP standard. By virtue of the high resolvability, we also show the ability to map the environment and thus operate as a bistatic sensing system.
We have detailed the hardware used for the real tests, including a commercial mmWave \ac{bs} and a vehicle-mounted cellular receiver combined with an RTK-enabled GNSS receiver as ground truth. We have also described the considered low-complexity channel estimation algorithms to process the received measurements, and proposed a \ac{bs} calibration algorithm, as well as positioning algorithms using only \ac{los} paths, and  a positioning and mapping algorithm using \ac{los} and \ac{nlos} paths. Via real-life data, we demonstrated that \ac{ue} positioning is feasible with a single BS. One of the main findings is that positioning based on multipath TDOA exploitation is sensitive to the environment geometry and may lead to poor positioning results. Positioning based on RTT, on the other hand, leads to good performance (below 2-meter error), provided AOA, AOD, and simulated RTT information is used. Correspondingly, environment mapping with RTT exhibits far better performance compared to based on multipath TDOA. Another important finding is that BS calibration is of high importance, not least in terms of the BS antenna panels, if downlink-AOD or uplink-AOA information is to be used. Such calibration is generally not needed for communication but is critical for positioning.
Overall, the results indicate that there is still a significant gap between theory and practice, which is ascribed to the combined effect of (i) BS calibration; (ii) synchronization errors; (iii) limited angular resolution in elevation, especially at the UE side; and (iv) limited knowledge of the used beam-patterns. If all these effects are mitigated, we foresee that sub-meter accuracy is attainable with a single BS. The inclusion of optimized signal design (precoding and combining) to boost the positioning accuracy, and the extension to a \ac{slam} problem where the UE motion model  is expected to further improve the positioning accuracy.

\appendices
\section{Geometric Relationships}\label{GeometricRelationships}
Within the geometric relationship $\boldsymbol{h}(\boldsymbol{s}_{k},\boldsymbol{s}_{\text{BS}},\boldsymbol{p}_{\text{IP},k}^{i} )$, the \ac{TOA} $\tau_{k}^{i}$ is determined by the propagation distance and the clock bias $b_{k}$, given by
\begin{align}
    \tau_{k}^{i} = \begin{cases} ||\boldsymbol{p}_{\text{BS}}-\boldsymbol{p}_{\text{UE},k}||/c + b_{k} & i=0 \\  ||\boldsymbol{p}_{\text{IP},k}^{i}-\boldsymbol{p}_{\text{UE},k}||/c + ||\boldsymbol{p}_{\text{IP},k}^{i}-\boldsymbol{p}_{\text{BS}}||/c +b_{k} & i\neq0\end{cases} \label{delay},
\end{align}
with $c$ denoting the speed of light, which is prior known. The \ac{AOA} $\boldsymbol{\theta}_{k}^{i}$ is determined by the arrival direction of the signal at the receiver $\boldsymbol{q}_{\text{AOA},k}^{i}$, and have components
\begin{align}
    &\theta_{\text{az},k}^{i} = \mathrm{arctan2}([\boldsymbol{q}_{\text{AOA},k}^{i}]_{2},[\boldsymbol{q}_{\text{AOA},k}^{i}]_{1}),\label{AOAAZ}\\
    &\theta_{\text{el},k}^{i} = \arcsin([\boldsymbol{q}_{\text{AOA},k}^{i}]_{3},||\boldsymbol{q}_{\text{AOA},k}^{i}||).\label{AOAEL}
\end{align}
The arrival direction of the signal $\boldsymbol{q}_{\text{AOA},k}^{i}$ is considered in the local coordinate system of the receiver so that it can be calculated by 
\begin{align}
 \boldsymbol{q}_{\text{AOA},k}^{i} = \begin{cases} \boldsymbol{R}_{\text{UE},k}(\boldsymbol{p}_{\text{BS}}-\boldsymbol{p}_{\text{UE},k}) & i=0 \\ \boldsymbol{R}_{\text{UE},k}(\boldsymbol{p}_{\text{IP},k}^{i}-\boldsymbol{p}_{\text{UE},k}) & i\neq0\end{cases}.
\end{align}
where $\boldsymbol{R}_{\text{UE},k}$ is the rotation matrix from the global reference coordinate system to the local coordinate system of the \ac{ue} \ac{URA} at time step $k$, determined by $\boldsymbol{\psi}_{\text{UE},k}$ as \cite{blanco2010tutorial}
\begin{align}
    &\boldsymbol{R}_{\text{UE},k} = \left[ \begin{matrix} \cos \gamma_{\text{UE},k} & -\sin \gamma_{\text{UE},k} & 0 \\ \sin \gamma_{\text{UE},k} &\cos \gamma_{\text{UE},k} & 0 \\ 0 & 0 & 1 \end{matrix}\right]  \label{eq:Euler2Rot} \left[ \begin{matrix} \cos \beta_{\text{UE},k} &0& \sin \beta_{\text{UE},k}  \\ 0 &1&0\\-\sin \beta_{\text{UE},k} &0&\cos \beta_{\text{UE},k}\end{matrix}\right] \left[ \begin{matrix} 1 &0 & 0 \\ 0 &\cos \alpha_{\text{UE},k} & -\sin \alpha_{\text{UE},k}  \\ 0 & \sin \alpha_{\text{UE},k} &\cos \alpha_{\text{UE},k}\end{matrix}\right], 
\end{align}
then, $\boldsymbol{R}_{\text{UE},k}^{\textsf{T}}$ denotes the rotation matrix from the local coordinate system to the global reference by definitely. Similarly, the \ac{AOD} $\boldsymbol{\phi}_{k}^{i}$ is determined by the arrival direction of the signal at the transmitter $\boldsymbol{q}_{\text{AOD},k}^{i}$, and have components
\begin{align}
    &\phi_{\text{az},k}^{i} = \mathrm{arctan2}([\boldsymbol{q}_{\text{AOD},k}^{i}]_{2},[\boldsymbol{q}_{\text{AOD},k}^{i}]_{1}),\label{AODAZ}\\
    &\phi_{\text{el},k}^{i} = \arcsin([\boldsymbol{q}_{\text{AOD},k}^{i}]_{3},||\boldsymbol{q}_{\text{AOD},k}^{i}||),\label{AODEL}
\end{align}
where $\boldsymbol{q}_{\text{AOA},k}^{i}$ is considered in the local coordinate system of the transmitter, given by
\begin{align}
    \boldsymbol{q}_{\text{AOD},k}^{i} = \begin{cases} \boldsymbol{R}_{\text{BS}}(\boldsymbol{p}_{\text{UE},k}-\boldsymbol{p}_{\text{BS}}) & i=0 \\ \boldsymbol{R}_{\text{BS}}(\boldsymbol{p}_{\text{IP},k}^{i}-\boldsymbol{p}_{\text{BS}}) & i\neq0\end{cases},
\end{align}
with $\boldsymbol{R}_{\text{BS}}$ denoting the rotation matrix from the global reference coordinate system to the local coordinate system of the \ac{bs} \ac{URA}, which is determined by $\boldsymbol{\psi}_{\text{BS}}$ as  % denoted by %\cite{blanco2010tutorial}
\begin{align}
    &\boldsymbol{R}_{\text{BS}} = \left[ \begin{matrix} \cos \gamma_{\text{BS}} & -\sin \gamma_{\text{BS}} & 0 \\ \sin \gamma_{\text{BS}} &\cos \gamma_{\text{BS}} & 0 \\ 0 & 0 & 1 \end{matrix}\right] \left[ \begin{matrix} \cos \beta_{\text{BS}} &0& \sin \beta_{\text{BS}}  \\ 0 &1&0\\-\sin \beta_{\text{BS}} &0&\cos \beta_{\text{BS}}\end{matrix}\right] \left[ \begin{matrix} 1 &0 & 0 \\ 0 &\cos \alpha_{\text{BS}} & -\sin \alpha_{\text{BS}}  \\ 0 & \sin \alpha_{\text{BS}} &\cos \alpha_{\text{BS}}\end{matrix}\right],
\end{align}
and $\boldsymbol{R}_{\text{BS}}^{\textsf{T}}$ is the rotation matrix from local coordinate system of the \ac{bs} to the global reference by definition.

\balance 
\bibliography{IEEEabrv,Bibliography}

% \IEEEtriggeratref{7}
% \IEEEtriggercmd{\enlargethispage{-20cm}}


%\bibliographystyle{IEEEtran}

\end{document}

%% file: acronyms.tex
\acrodef{bs}[BS]{base station}
\acrodef{ue}[UE]{user equipment}
\acrodef{URA}[URA]{uniform rectangular array}
\acrodef{VA}[VA]{virtual anchor}
\acrodef{TOA}[TOA]{time-of-arrival}
\acrodef{rtt}[RTT]{round-trip-time}
\acrodef{TDOA}[TDOA]{time-difference-of-arrival}
\acrodef{AOA}[AOA]{angles-of-arrival}
\acrodef{AOD}[AOD]{angles-of-departure}
\acrodef{los}[LOS]{line-of-sight}
\acrodef{nlos}[NLOS]{non-line-of-sight}
\acrodef{ad}[AD]{autonomous drive}
\acrodef{adas}[ADAS]{advanced driver assistance system}
\acrodef{gnss}[GNSS]{global navigation satellite system}
\acrodef{imu}[IMU]{inertial measurement unit}
\acrodef{prs}[PRS]{positioning reference signal}
\acrodef{MAP}[MAP]{maximum a posteriori}
\acrodef{ML}[ML]{maximum likelihood}
\acrodef{MC}[MC]{Monte Carlo}
\acrodef{AAS}[AAS]{active antenna system}
\acrodef{SSB}[SSB]{synchronization signal/physical broadcast channel block}
\acrodef{CSI-RS}[CSI-RS]{channel state information reference signal}
\acrodef{PSS}[PSS]{primary synchronization signal}
\acrodef{SSS}[SSS]{secondary synchronization signal}
\acrodef{PBCH}[PBCH]{physical broadcast channel}
\acrodef{PBCH-DMRS}[PBCH-DMRS]{physical broadcast channel-demodulation reference signal}
\acrodef{RF}[RF]{radio frequency}
\acrodef{MAE}[MAE]{mean absolute value}

\acrodef{RTK}[RTK]{real-time kinematic}
\acrodef{GNSS}[GNSS]{global navigation satellite system}
\acrodef{IMU}[IMU]{inertial measurement unit}

\acrodef{RFSoC}[RFSoC]{radio frequency system-on-chip}
\acrodef{PPS}[PPS]{pulse per second}
\acrodef{OFDM}[OFDM]{orthogonal frequency division multiplexing}
\acrodef{GPS}[GPS]{global positioning system}
\acrodef{MPC}[MPC]{multipath component}
\acrodef{DBSCAN}[DBSCAN]{density-based spatial clustering of applications with noise}

\acrodef{CDF}[CDF]{cumulative distribution function}
\acrodef{slam}[SLAM]{simultaneous localization and mapping}

\acrodef{IP}[IP]{incidence point}
\acrodef{RIS}[RIS]{reconfigurable intelligent surface}
\acrodef{RSS}[RSS]{received signal strength}
\acrodef{AD}[AD]{autonomous drive}
\acrodef{ADAS}[ADAS]{advanced driving assistance service}

%% file: Figures_without_ESPRIT/c_1.tex
\begin{tikzpicture}

\begin{axis}[%
width=6.3cm,
height=4cm,
scale only axis,
scale only axis,
point meta min=80,
point meta max=115,
axis on top,
xmin=0.5,
xmax=34.5,
xlabel style={font=\color{white!15!black}},
xlabel={BS Beam Index ($-57^\circ$ to $57^\circ$)},
ymin=0.5,
ymax=15.5,
ylabel style={font=\color{white!15!black}},
ylabel={UE Beam Index ($-45^\circ$ to $45^\circ$)},
axis background/.style={fill=white},
colormap={mymap}{[1pt] rgb(0pt)=(0.2422,0.1504,0.6603); rgb(1pt)=(0.2444,0.1534,0.6728); rgb(2pt)=(0.2464,0.1569,0.6847); rgb(3pt)=(0.2484,0.1607,0.6961); rgb(4pt)=(0.2503,0.1648,0.7071); rgb(5pt)=(0.2522,0.1689,0.7179); rgb(6pt)=(0.254,0.1732,0.7286); rgb(7pt)=(0.2558,0.1773,0.7393); rgb(8pt)=(0.2576,0.1814,0.7501); rgb(9pt)=(0.2594,0.1854,0.761); rgb(11pt)=(0.2628,0.1932,0.7828); rgb(12pt)=(0.2645,0.1972,0.7937); rgb(13pt)=(0.2661,0.2011,0.8043); rgb(14pt)=(0.2676,0.2052,0.8148); rgb(15pt)=(0.2691,0.2094,0.8249); rgb(16pt)=(0.2704,0.2138,0.8346); rgb(17pt)=(0.2717,0.2184,0.8439); rgb(18pt)=(0.2729,0.2231,0.8528); rgb(19pt)=(0.274,0.228,0.8612); rgb(20pt)=(0.2749,0.233,0.8692); rgb(21pt)=(0.2758,0.2382,0.8767); rgb(22pt)=(0.2766,0.2435,0.884); rgb(23pt)=(0.2774,0.2489,0.8908); rgb(24pt)=(0.2781,0.2543,0.8973); rgb(25pt)=(0.2788,0.2598,0.9035); rgb(26pt)=(0.2794,0.2653,0.9094); rgb(27pt)=(0.2798,0.2708,0.915); rgb(28pt)=(0.2802,0.2764,0.9204); rgb(29pt)=(0.2806,0.2819,0.9255); rgb(30pt)=(0.2809,0.2875,0.9305); rgb(31pt)=(0.2811,0.293,0.9352); rgb(32pt)=(0.2813,0.2985,0.9397); rgb(33pt)=(0.2814,0.304,0.9441); rgb(34pt)=(0.2814,0.3095,0.9483); rgb(35pt)=(0.2813,0.315,0.9524); rgb(36pt)=(0.2811,0.3204,0.9563); rgb(37pt)=(0.2809,0.3259,0.96); rgb(38pt)=(0.2807,0.3313,0.9636); rgb(39pt)=(0.2803,0.3367,0.967); rgb(40pt)=(0.2798,0.3421,0.9702); rgb(41pt)=(0.2791,0.3475,0.9733); rgb(42pt)=(0.2784,0.3529,0.9763); rgb(43pt)=(0.2776,0.3583,0.9791); rgb(44pt)=(0.2766,0.3638,0.9817); rgb(45pt)=(0.2754,0.3693,0.984); rgb(46pt)=(0.2741,0.3748,0.9862); rgb(47pt)=(0.2726,0.3804,0.9881); rgb(48pt)=(0.271,0.386,0.9898); rgb(49pt)=(0.2691,0.3916,0.9912); rgb(50pt)=(0.267,0.3973,0.9924); rgb(51pt)=(0.2647,0.403,0.9935); rgb(52pt)=(0.2621,0.4088,0.9946); rgb(53pt)=(0.2591,0.4145,0.9955); rgb(54pt)=(0.2556,0.4203,0.9965); rgb(55pt)=(0.2517,0.4261,0.9974); rgb(56pt)=(0.2473,0.4319,0.9983); rgb(57pt)=(0.2424,0.4378,0.9991); rgb(58pt)=(0.2369,0.4437,0.9996); rgb(59pt)=(0.2311,0.4497,0.9995); rgb(60pt)=(0.225,0.4559,0.9985); rgb(61pt)=(0.2189,0.462,0.9968); rgb(62pt)=(0.2128,0.4682,0.9948); rgb(63pt)=(0.2066,0.4743,0.9926); rgb(64pt)=(0.2006,0.4803,0.9906); rgb(65pt)=(0.195,0.4861,0.9887); rgb(66pt)=(0.1903,0.4919,0.9867); rgb(67pt)=(0.1869,0.4975,0.9844); rgb(68pt)=(0.1847,0.503,0.9819); rgb(69pt)=(0.1831,0.5084,0.9793); rgb(70pt)=(0.1818,0.5138,0.9766); rgb(71pt)=(0.1806,0.5191,0.9738); rgb(72pt)=(0.1795,0.5244,0.9709); rgb(73pt)=(0.1785,0.5296,0.9677); rgb(74pt)=(0.1778,0.5349,0.9641); rgb(75pt)=(0.1773,0.5401,0.9602); rgb(76pt)=(0.1768,0.5452,0.956); rgb(77pt)=(0.1764,0.5504,0.9516); rgb(78pt)=(0.1755,0.5554,0.9473); rgb(79pt)=(0.174,0.5605,0.9432); rgb(80pt)=(0.1716,0.5655,0.9393); rgb(81pt)=(0.1686,0.5705,0.9357); rgb(82pt)=(0.1649,0.5755,0.9323); rgb(83pt)=(0.161,0.5805,0.9289); rgb(84pt)=(0.1573,0.5854,0.9254); rgb(85pt)=(0.154,0.5902,0.9218); rgb(86pt)=(0.1513,0.595,0.9182); rgb(87pt)=(0.1492,0.5997,0.9147); rgb(88pt)=(0.1475,0.6043,0.9113); rgb(89pt)=(0.1461,0.6089,0.908); rgb(90pt)=(0.1446,0.6135,0.905); rgb(91pt)=(0.1429,0.618,0.9022); rgb(92pt)=(0.1408,0.6226,0.8998); rgb(93pt)=(0.1383,0.6272,0.8975); rgb(94pt)=(0.1354,0.6317,0.8953); rgb(95pt)=(0.1321,0.6363,0.8932); rgb(96pt)=(0.1288,0.6408,0.891); rgb(97pt)=(0.1253,0.6453,0.8887); rgb(98pt)=(0.1219,0.6497,0.8862); rgb(99pt)=(0.1185,0.6541,0.8834); rgb(100pt)=(0.1152,0.6584,0.8804); rgb(101pt)=(0.1119,0.6627,0.877); rgb(102pt)=(0.1085,0.6669,0.8734); rgb(103pt)=(0.1048,0.671,0.8695); rgb(104pt)=(0.1009,0.675,0.8653); rgb(105pt)=(0.0964,0.6789,0.8609); rgb(106pt)=(0.0914,0.6828,0.8562); rgb(107pt)=(0.0855,0.6865,0.8513); rgb(108pt)=(0.0789,0.6902,0.8462); rgb(109pt)=(0.0713,0.6938,0.8409); rgb(110pt)=(0.0628,0.6972,0.8355); rgb(111pt)=(0.0535,0.7006,0.8299); rgb(112pt)=(0.0433,0.7039,0.8242); rgb(113pt)=(0.0328,0.7071,0.8183); rgb(114pt)=(0.0234,0.7103,0.8124); rgb(115pt)=(0.0155,0.7133,0.8064); rgb(116pt)=(0.0091,0.7163,0.8003); rgb(117pt)=(0.0046,0.7192,0.7941); rgb(118pt)=(0.0019,0.722,0.7878); rgb(119pt)=(0.0009,0.7248,0.7815); rgb(120pt)=(0.0018,0.7275,0.7752); rgb(121pt)=(0.0046,0.7301,0.7688); rgb(122pt)=(0.0094,0.7327,0.7623); rgb(123pt)=(0.0162,0.7352,0.7558); rgb(124pt)=(0.0253,0.7376,0.7492); rgb(125pt)=(0.0369,0.74,0.7426); rgb(126pt)=(0.0504,0.7423,0.7359); rgb(127pt)=(0.0638,0.7446,0.7292); rgb(128pt)=(0.077,0.7468,0.7224); rgb(129pt)=(0.0899,0.7489,0.7156); rgb(130pt)=(0.1023,0.751,0.7088); rgb(131pt)=(0.1141,0.7531,0.7019); rgb(132pt)=(0.1252,0.7552,0.695); rgb(133pt)=(0.1354,0.7572,0.6881); rgb(134pt)=(0.1448,0.7593,0.6812); rgb(135pt)=(0.1532,0.7614,0.6741); rgb(136pt)=(0.1609,0.7635,0.6671); rgb(137pt)=(0.1678,0.7656,0.6599); rgb(138pt)=(0.1741,0.7678,0.6527); rgb(139pt)=(0.1799,0.7699,0.6454); rgb(140pt)=(0.1853,0.7721,0.6379); rgb(141pt)=(0.1905,0.7743,0.6303); rgb(142pt)=(0.1954,0.7765,0.6225); rgb(143pt)=(0.2003,0.7787,0.6146); rgb(144pt)=(0.2061,0.7808,0.6065); rgb(145pt)=(0.2118,0.7828,0.5983); rgb(146pt)=(0.2178,0.7849,0.5899); rgb(147pt)=(0.2244,0.7869,0.5813); rgb(148pt)=(0.2318,0.7887,0.5725); rgb(149pt)=(0.2401,0.7905,0.5636); rgb(150pt)=(0.2491,0.7922,0.5546); rgb(151pt)=(0.2589,0.7937,0.5454); rgb(152pt)=(0.2695,0.7951,0.536); rgb(153pt)=(0.2809,0.7964,0.5266); rgb(154pt)=(0.2929,0.7975,0.517); rgb(155pt)=(0.3052,0.7985,0.5074); rgb(156pt)=(0.3176,0.7994,0.4975); rgb(157pt)=(0.3301,0.8002,0.4876); rgb(158pt)=(0.3424,0.8009,0.4774); rgb(159pt)=(0.3548,0.8016,0.4669); rgb(160pt)=(0.3671,0.8021,0.4563); rgb(161pt)=(0.3795,0.8026,0.4454); rgb(162pt)=(0.3921,0.8029,0.4344); rgb(163pt)=(0.405,0.8031,0.4233); rgb(164pt)=(0.4184,0.803,0.4122); rgb(165pt)=(0.4322,0.8028,0.4013); rgb(166pt)=(0.4463,0.8024,0.3904); rgb(167pt)=(0.4608,0.8018,0.3797); rgb(168pt)=(0.4753,0.8011,0.3691); rgb(169pt)=(0.4899,0.8002,0.3586); rgb(170pt)=(0.5044,0.7993,0.348); rgb(171pt)=(0.5187,0.7982,0.3374); rgb(172pt)=(0.5329,0.797,0.3267); rgb(173pt)=(0.547,0.7957,0.3159); rgb(175pt)=(0.5748,0.7929,0.2941); rgb(176pt)=(0.5886,0.7913,0.2833); rgb(177pt)=(0.6024,0.7896,0.2726); rgb(178pt)=(0.6161,0.7878,0.2622); rgb(179pt)=(0.6297,0.7859,0.2521); rgb(180pt)=(0.6433,0.7839,0.2423); rgb(181pt)=(0.6567,0.7818,0.2329); rgb(182pt)=(0.6701,0.7796,0.2239); rgb(183pt)=(0.6833,0.7773,0.2155); rgb(184pt)=(0.6963,0.775,0.2075); rgb(185pt)=(0.7091,0.7727,0.1998); rgb(186pt)=(0.7218,0.7703,0.1924); rgb(187pt)=(0.7344,0.7679,0.1852); rgb(188pt)=(0.7468,0.7654,0.1782); rgb(189pt)=(0.759,0.7629,0.1717); rgb(190pt)=(0.771,0.7604,0.1658); rgb(191pt)=(0.7829,0.7579,0.1608); rgb(192pt)=(0.7945,0.7554,0.157); rgb(193pt)=(0.806,0.7529,0.1546); rgb(194pt)=(0.8172,0.7505,0.1535); rgb(195pt)=(0.8281,0.7481,0.1536); rgb(196pt)=(0.8389,0.7457,0.1546); rgb(197pt)=(0.8495,0.7435,0.1564); rgb(198pt)=(0.86,0.7413,0.1587); rgb(199pt)=(0.8703,0.7392,0.1615); rgb(200pt)=(0.8804,0.7372,0.165); rgb(201pt)=(0.8903,0.7353,0.1695); rgb(202pt)=(0.9,0.7336,0.1749); rgb(203pt)=(0.9093,0.7321,0.1815); rgb(204pt)=(0.9184,0.7308,0.189); rgb(205pt)=(0.9272,0.7298,0.1973); rgb(206pt)=(0.9357,0.729,0.2061); rgb(207pt)=(0.944,0.7285,0.2151); rgb(208pt)=(0.9523,0.7284,0.2237); rgb(209pt)=(0.9606,0.7285,0.2312); rgb(210pt)=(0.9689,0.7292,0.2373); rgb(211pt)=(0.977,0.7304,0.2418); rgb(212pt)=(0.9842,0.733,0.2446); rgb(213pt)=(0.99,0.7365,0.2429); rgb(214pt)=(0.9946,0.7407,0.2394); rgb(215pt)=(0.9966,0.7458,0.2351); rgb(216pt)=(0.9971,0.7513,0.2309); rgb(217pt)=(0.9972,0.7569,0.2267); rgb(218pt)=(0.9971,0.7626,0.2224); rgb(219pt)=(0.9969,0.7683,0.2181); rgb(220pt)=(0.9966,0.774,0.2138); rgb(221pt)=(0.9962,0.7798,0.2095); rgb(222pt)=(0.9957,0.7856,0.2053); rgb(223pt)=(0.9949,0.7915,0.2012); rgb(224pt)=(0.9938,0.7974,0.1974); rgb(225pt)=(0.9923,0.8034,0.1939); rgb(226pt)=(0.9906,0.8095,0.1906); rgb(227pt)=(0.9885,0.8156,0.1875); rgb(228pt)=(0.9861,0.8218,0.1846); rgb(229pt)=(0.9835,0.828,0.1817); rgb(230pt)=(0.9807,0.8342,0.1787); rgb(231pt)=(0.9778,0.8404,0.1757); rgb(232pt)=(0.9748,0.8467,0.1726); rgb(233pt)=(0.972,0.8529,0.1695); rgb(234pt)=(0.9694,0.8591,0.1665); rgb(235pt)=(0.9671,0.8654,0.1636); rgb(236pt)=(0.9651,0.8716,0.1608); rgb(237pt)=(0.9634,0.8778,0.1582); rgb(238pt)=(0.9619,0.884,0.1557); rgb(239pt)=(0.9608,0.8902,0.1532); rgb(240pt)=(0.9601,0.8963,0.1507); rgb(241pt)=(0.9596,0.9023,0.148); rgb(242pt)=(0.9595,0.9084,0.145); rgb(243pt)=(0.9597,0.9143,0.1418); rgb(244pt)=(0.9601,0.9203,0.1382); rgb(245pt)=(0.9608,0.9262,0.1344); rgb(246pt)=(0.9618,0.932,0.1304); rgb(247pt)=(0.9629,0.9379,0.1261); rgb(248pt)=(0.9642,0.9437,0.1216); rgb(249pt)=(0.9657,0.9494,0.1168); rgb(250pt)=(0.9674,0.9552,0.1116); rgb(251pt)=(0.9692,0.9609,0.1061); rgb(252pt)=(0.9711,0.9667,0.1001); rgb(253pt)=(0.973,0.9724,0.0938); rgb(254pt)=(0.9749,0.9782,0.0872); rgb(255pt)=(0.9769,0.9839,0.0805)},
colorbar
]
\addplot [forget plot] graphics [xmin=0, xmax=35, ymin=0, ymax=16] {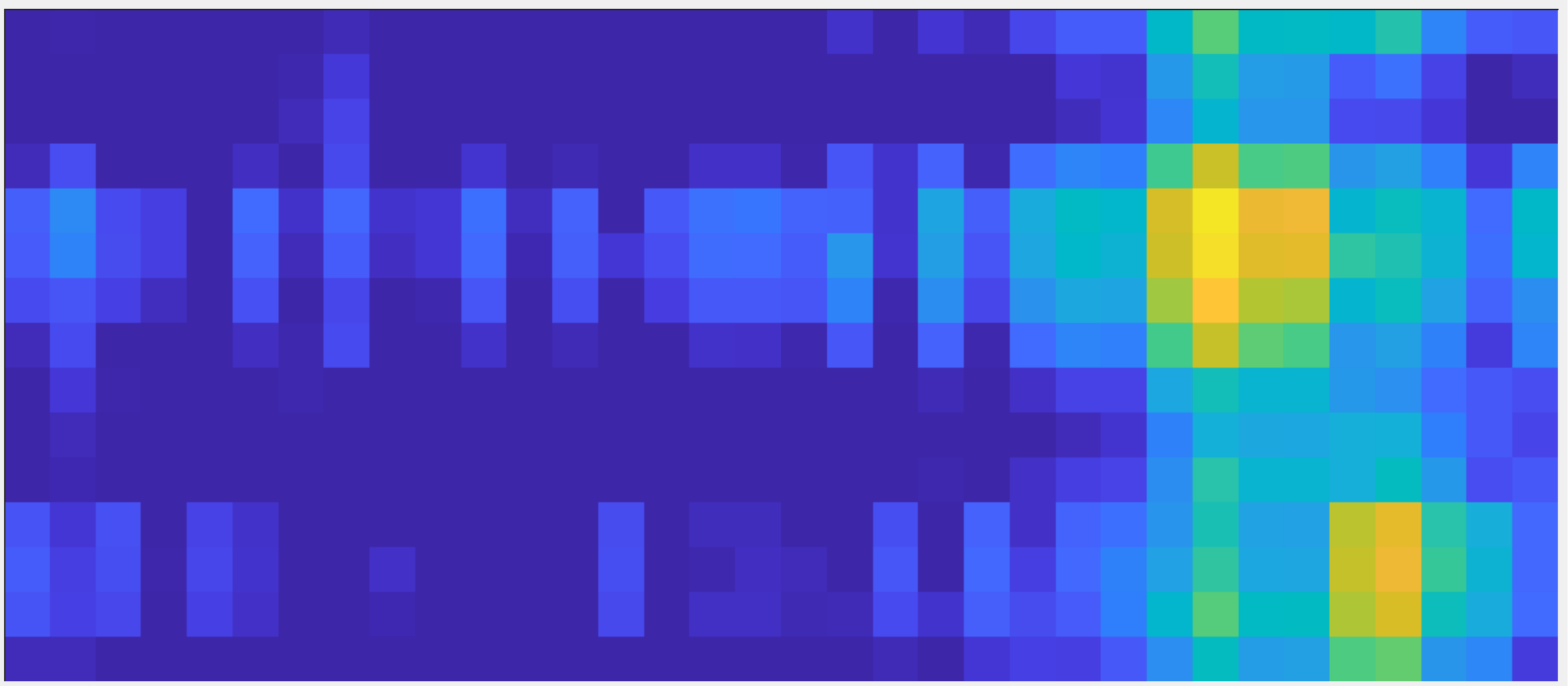};
\end{axis}

\end{tikzpicture}%

%% file: Figures_without_ESPRIT/c_2_3times.tex
\definecolor{mycolor1}{rgb}{0.00000,0.44700,0.74100}%
\begin{tikzpicture}

\begin{axis}[%
width=7.2 cm,
height=4. cm,
scale only axis,
xmin=0,
xmax=350,
xlabel style={font=\color{white!15!black},font=\normalsize},
xlabel={Measurement Index},
ymin=-200,
ymax=150,
ylabel style={font=\color{white!15!black},font=\normalsize},
ylabel={Delay [m]},
axis background/.style={fill=white},
axis x line*=bottom,
axis y line*=left,
xmajorgrids,
ymajorgrids,
%xmajorgrids,
%ymajorgrids,
%yminorgrids，
legend columns=3, 
legend style={font=\scriptsize, at={(0, 0)}, anchor=south west, legend cell align=left, align=left, draw=white!15!black}
]
\addplot [color=green, line width=1.0pt]
  table[row sep=crcr]{%
1	117.312825630685\\
2	117.30515586841\\
3	117.252252510966\\
4	116.997086379603\\
5	116.68943796516\\
6	116.394386847921\\
7	116.082987924317\\
8	115.779731618308\\
9	115.622248761256\\
10	115.498507055906\\
11	115.357816194616\\
12	115.206536085247\\
13	115.08377797638\\
14	114.989693641512\\
15	114.936336706412\\
16	114.96332435487\\
17	114.953538113454\\
18	114.990268137666\\
19	115.031134950709\\
20	115.097697035576\\
21	115.193803187574\\
22	115.207884710996\\
23	115.180173077418\\
24	115.235991941646\\
25	115.360613974334\\
26	115.550406212174\\
27	115.879426658825\\
28	116.095280126431\\
29	116.359364989987\\
30	116.824251188129\\
31	117.162553701806\\
32	117.542512756719\\
33	117.931596478597\\
34	118.421434548711\\
35	118.917326696259\\
36	119.600900312058\\
37	120.250984083924\\
38	120.792017758183\\
39	121.334045590281\\
40	122.014182606765\\
41	122.565165700264\\
42	123.044199734329\\
43	123.834358334395\\
44	124.546430181595\\
45	125.380761048806\\
46	125.936270790518\\
47	126.567620220393\\
48	127.111890353559\\
49	127.729606388465\\
50	128.294133948851\\
51	128.878126241906\\
52	129.332042712594\\
53	129.587909548722\\
54	129.484149957522\\
55	128.831528323811\\
56	127.036941873111\\
57	124.615259632679\\
58	122.503118676548\\
59	121.413274832967\\
60	120.298613575355\\
61	119.583660682724\\
62	118.620625878651\\
63	117.417052925043\\
64	116.304664310963\\
65	115.318799236581\\
66	114.304890622382\\
67	113.440058529155\\
68	112.659596035764\\
69	111.698900219903\\
70	110.733659222466\\
71	109.790367960512\\
72	108.513851722323\\
73	107.681815463296\\
74	106.630622228819\\
75	105.34806219713\\
76	104.080475193328\\
77	102.989956736295\\
78	102.107479715289\\
79	101.3035073985\\
80	100.326897855445\\
81	98.8863215647904\\
82	97.818998681192\\
83	97.0404509598006\\
84	95.8218575887324\\
85	94.503353470089\\
86	93.0882740517184\\
87	92.0891169777895\\
88	90.4464547496358\\
89	88.9665379441257\\
90	87.689910420911\\
91	85.9227462948194\\
92	84.1983875904822\\
93	82.9605605827503\\
94	81.7718925849052\\
95	80.6456024717415\\
96	79.4868018911456\\
97	77.9919002672665\\
98	76.4113503867824\\
99	75.4164240589208\\
100	74.2716220997416\\
101	73.004288101967\\
102	71.9442304734391\\
103	71.3521202452726\\
104	70.2540457452342\\
105	69.0008725092942\\
106	68.0003746868373\\
107	67.1761225325026\\
108	66.0929435863694\\
109	65.0237603939135\\
110	64.2906997067377\\
111	63.9028387979594\\
112	63.5888596001299\\
113	63.1150157719412\\
114	62.5596988482779\\
115	62.1831784450329\\
116	61.8377937161824\\
117	61.4533697262275\\
118	61.1217862660116\\
119	60.8620191356472\\
120	60.643078018504\\
121	60.5846685596209\\
122	60.6670026908501\\
123	116.77632828531\\
124	116.468528812611\\
125	116.181130114945\\
126	115.908185732426\\
127	115.702172250353\\
128	115.500086769201\\
129	115.312532742158\\
130	115.148892829155\\
131	114.987997676077\\
132	114.827804292974\\
133	114.790458301941\\
134	114.721574617698\\
135	114.691234690915\\
136	114.686675820783\\
137	114.723599534002\\
138	114.734474507203\\
139	114.791838405984\\
140	114.885862956807\\
141	114.936606982915\\
142	114.998121492172\\
143	115.242079140655\\
144	115.409404671757\\
145	115.679781403103\\
146	116.042636642752\\
147	116.300957864915\\
148	116.730255629469\\
149	117.121885121723\\
150	117.518964442579\\
151	117.955121004288\\
152	118.453365856448\\
153	118.964573925799\\
154	119.55961919532\\
155	120.252140368201\\
156	120.802863948076\\
157	121.468978946526\\
158	122.259818283491\\
159	123.098802433213\\
160	123.766837321833\\
161	124.5830266682\\
162	125.395573528604\\
163	126.255654212306\\
164	127.157411147376\\
165	127.847263300597\\
166	128.79781372501\\
167	129.465516884426\\
168	129.932373931787\\
169	130.037908337758\\
170	129.75258185352\\
171	129.035408718681\\
172	127.994978366137\\
173	126.551749868449\\
174	125.177298121926\\
175	123.877442286015\\
176	122.272236915156\\
177	120.789312476816\\
178	119.274094677779\\
179	118.119742035129\\
180	116.938799967935\\
181	115.744763244668\\
182	114.47736779199\\
183	113.133142839894\\
184	111.803233310871\\
185	110.48935647884\\
186	109.178286276536\\
187	107.877962070165\\
188	106.267369808663\\
189	104.89623730779\\
190	103.463204014532\\
191	101.974805658504\\
192	100.446638133688\\
193	98.8645386266995\\
194	96.6170631309359\\
195	95.3202628960562\\
196	94.3139318267649\\
197	92.9338656303562\\
198	91.6721118184649\\
199	90.7436438200306\\
200	89.9351572158002\\
201	88.7346907833171\\
202	87.502029613843\\
203	86.5708843289453\\
204	85.5835595834326\\
205	84.4894981916113\\
206	83.2034150504899\\
207	82.3804581743599\\
208	81.3799811352804\\
209	79.9885472225027\\
210	78.5491117272247\\
211	76.992188059119\\
212	75.6644428194967\\
213	74.094285417659\\
214	72.5028327751431\\
215	71.0530406260682\\
216	69.7379725772814\\
217	68.8539708406648\\
218	67.9276632235406\\
219	66.9483508100245\\
220	66.0724521911765\\
221	65.0796842550303\\
222	64.164865863221\\
223	63.1516989204485\\
224	62.5449915832658\\
225	62.0175480458084\\
226	61.470184071898\\
227	60.9915679195261\\
228	60.7965142612692\\
229	60.5852355340841\\
230	60.2631703112476\\
231	60.1750293987831\\
232	60.1012442919722\\
233	59.987578921218\\
234	59.9585320068428\\
235	59.9464007691774\\
236	59.9887314605097\\
237	60.0392601144516\\
238	60.1211268942277\\
239	60.3134127535342\\
240	114.372349969438\\
241	113.903169743103\\
242	113.53033327348\\
243	113.322979828262\\
244	113.167529121472\\
245	113.076952472033\\
246	113.043273332109\\
247	112.977627276784\\
248	112.916308075312\\
249	112.888002309664\\
250	112.865263367497\\
251	112.856522371637\\
252	112.862442320421\\
253	112.943362872266\\
254	113.05454429196\\
255	113.287791981647\\
256	113.593934244953\\
257	113.977098189502\\
258	114.535977985693\\
259	115.012415979413\\
260	115.509092545971\\
261	115.950184074811\\
262	115.967074175339\\
263	115.960628046437\\
264	115.960665276972\\
265	116.163025190921\\
266	116.859719714343\\
267	117.504680400398\\
268	118.051585642114\\
269	118.699355162385\\
270	119.475268425383\\
271	120.338801658133\\
272	121.318049783649\\
273	122.711258901664\\
274	123.966014092755\\
275	124.822722898644\\
276	125.905400011112\\
277	126.783598542408\\
278	127.631032358101\\
279	128.275899402492\\
280	128.653098034556\\
281	128.854573489713\\
282	128.728374795787\\
283	128.167067381582\\
284	127.264681163221\\
285	125.805450916526\\
286	124.230560799919\\
287	123.175419126966\\
288	121.896102095395\\
289	120.361985449631\\
290	119.143655021632\\
291	117.683364716004\\
292	116.148592986517\\
293	115.19111321877\\
294	113.882216548212\\
295	112.319584948944\\
296	110.910698360371\\
297	109.61281689286\\
298	108.556092954998\\
299	107.328978281076\\
300	105.822087224446\\
301	104.069179571665\\
302	103.079581119358\\
303	101.5483012189\\
304	99.9483829884341\\
305	98.5424181816865\\
306	97.3757768859118\\
307	96.4505791133521\\
308	95.0857553793398\\
309	93.9444344863801\\
310	92.6481176729974\\
311	91.5916639186638\\
312	90.4722923618024\\
313	89.122659797657\\
314	87.4720821706801\\
315	85.9723026957374\\
316	84.786020431978\\
317	84.1388359831558\\
318	83.3178958783894\\
319	82.3309493910957\\
320	81.5028421237169\\
321	80.3433797041884\\
322	79.0212023535056\\
323	77.6751794798795\\
324	76.6206928555506\\
325	75.4674045993107\\
326	74.1555005006075\\
327	72.9707092082423\\
328	71.8084743755045\\
329	70.6616370624725\\
330	69.6473715741692\\
331	68.6875092032365\\
332	67.7078643006065\\
333	66.9411174741656\\
334	66.0527479840132\\
335	65.0574505908283\\
336	63.9602149656394\\
337	63.0718740717499\\
338	62.3524590778339\\
339	61.6439731628541\\
340	61.0836651192763\\
341	60.7078583850153\\
342	60.4240148193293\\
343	60.3449903709296\\
344	60.4076318177114\\
345	60.5894552551675\\
346	60.8759033200762\\
347	61.2059797700258\\
};
\addlegendentry{Ground Truth}

\addplot [color=blue, only marks, mark=o, mark size=1pt, mark options={solid, blue}]
  table[row sep=crcr]{%
1	-87.3402434497988\\
2	-85.4443351443458\\
3	-86.5429580856056\\
4	-87.3915495754591\\
5	-87.5214724277734\\
6	-88.3952865376595\\
7	-88.9227004635169\\
8	-88.2494657642384\\
9	-88.6160571682513\\
10	-87.9019148500603\\
11	-86.4124722507408\\
12	-86.4034582585299\\
13	-85.0336654431298\\
14	-86.0020428429176\\
15	-82.8056783963644\\
16	-87.7417835384907\\
17	-86.52847991502\\
18	6.06103486955759\\
19	5.87748259658854\\
20	-88.2737172512256\\
21	-87.1952609773005\\
22	82.2498154579758\\
23	81.8357025125406\\
24	81.5756596980675\\
25	80.1896716572368\\
26	81.3809152140962\\
27	75.7931065389444\\
28	80.1794451599937\\
29	33.8388520324005\\
30	79.5592399952177\\
31	79.1280098833324\\
32	78.261683468021\\
33	56.7114614133935\\
34	80.2517788258\\
35	76.4048621584501\\
36	-82.2419155523166\\
37	72.6193786060365\\
38	-76.1921118315488\\
39	-75.5689776477711\\
40	75.76187977717\\
41	37.8741120282633\\
42	-78.5137917612087\\
43	23.1075563211809\\
44	52.2536955213332\\
45	69.7518559791463\\
46	48.8287463621905\\
47	-64.1023452303007\\
48	-70.9581640243171\\
49	-62.4453081102772\\
50	-48.129669238105\\
51	-65.2404339526306\\
52	65.5505436372599\\
53	63.8333575136313\\
54	-75.1382056107249\\
55	-75.2105844733942\\
56	-78.3256466945191\\
57	-79.3360111997994\\
58	-83.0143215829797\\
59	-82.6891753424133\\
60	-80.7817335487868\\
61	-82.6247134143688\\
62	-84.0245114541251\\
63	-84.2356898884094\\
64	-87.9931531205766\\
65	-87.6933777052104\\
66	-88.8720853325539\\
67	-91.8535771175577\\
68	-88.0147274586877\\
69	-92.107674057242\\
70	-93.6067293237867\\
71	-94.4101185863325\\
72	-88.3265328589683\\
73	-95.7906726888835\\
74	-85.4606047844513\\
75	-94.5093563268082\\
76	-97.6528563908251\\
77	-100.504838242869\\
78	-102.228636716752\\
79	-101.389705866327\\
80	-102.251604814203\\
81	-102.539344975837\\
82	-103.819266566901\\
83	-106.581142989659\\
84	-76.6177425102775\\
85	-109.826229548152\\
86	-109.223474495476\\
87	-110.879600133553\\
88	-102.83485175533\\
89	-88.1718256308382\\
90	-88.5577024991952\\
91	-68.4897726817819\\
92	-14.6249721896801\\
93	-120.006431409958\\
94	-122.303967731449\\
95	-122.949191980316\\
96	-124.390321405263\\
97	-125.985316294807\\
98	-126.175818911094\\
99	-127.421463359757\\
100	-129.881299908777\\
101	-131.957518314538\\
102	-129.634668239165\\
103	-131.174877117129\\
104	-131.154707414728\\
105	119.867586376731\\
106	-134.321876419378\\
107	-131.271586733995\\
108	166.905823175472\\
109	-135.99132920159\\
110	-58.1937840837623\\
111	-48.559234628795\\
112	-51.3473044785405\\
113	-53.5251726782446\\
114	-54.4047763762933\\
115	-55.1522246299921\\
116	-53.5447175705701\\
117	-61.0527094360969\\
118	-60.7240871719097\\
119	186.096543251364\\
120	-65.5421742082379\\
121	-62.8759429475148\\
122	-65.6798942959999\\
123	-88.2077119215507\\
124	-86.2360649399327\\
125	-86.6828923270131\\
126	-86.0234866551997\\
127	-85.6459086144277\\
128	-87.0880739539326\\
129	-86.8570144194279\\
130	-86.3882662416151\\
131	-89.8536332970792\\
132	-86.1479387809774\\
133	-86.9946605335355\\
134	-89.3473018741452\\
135	-90.0002982271649\\
136	-88.4714993011464\\
137	-87.9602669643767\\
138	-88.0204963513452\\
139	-86.811824461609\\
140	82.267003028119\\
141	82.3687797601915\\
142	-86.5919248144182\\
143	81.4851699012904\\
144	79.3282664590751\\
145	73.3013605859289\\
146	78.2488263167202\\
147	79.8712434132904\\
148	-80.5776328407594\\
149	75.2751444659267\\
150	76.6148589010326\\
151	76.4005682445041\\
152	76.9357871894863\\
153	73.4306032644715\\
154	27.6418855161265\\
155	72.0519111856784\\
156	72.3766160205927\\
157	-76.0778666450481\\
158	35.4130330128234\\
159	46.8890230838191\\
160	-74.1211857004355\\
161	-8.94877755471683\\
162	67.8242663438614\\
163	-67.9018342729147\\
164	-66.8239811429523\\
165	-63.535666435406\\
166	-61.1099095828147\\
167	-57.7130647936646\\
168	58.4914413802839\\
169	27.215190897334\\
170	-78.6739644298668\\
171	-79.3678529819816\\
172	-80.3541600882012\\
173	-78.6857737398001\\
174	-81.0182221158793\\
175	-82.0129952061288\\
176	-83.9078181881071\\
177	-82.8582560672512\\
178	-84.7747952348602\\
179	-86.5567406910282\\
180	-89.3296866357474\\
181	-90.9220133651231\\
182	-91.3699614987465\\
183	-93.5646819669836\\
184	-87.6638417853209\\
185	-93.6509678099755\\
186	-93.9133937797906\\
187	-96.9427774602445\\
188	-98.2960950072316\\
189	-100.897678604199\\
190	-102.963868703574\\
191	-101.512314113134\\
192	-105.029786376558\\
193	-107.599137046326\\
194	-111.386660867274\\
195	-112.368312217545\\
196	-114.108500267989\\
197	-114.403746922435\\
198	-104.765917674275\\
199	-49.6854990093812\\
200	-4.19929516568479\\
201	-107.379823676926\\
202	-97.3625628400749\\
203	-119.601928823404\\
204	-64.7430176504009\\
205	-83.6482094188595\\
206	-119.544524456741\\
207	-124.939384233984\\
208	-126.512392676287\\
209	-122.539491071382\\
210	-126.039671674361\\
211	-128.172416591136\\
212	-130.035950993702\\
213	-131.847666157718\\
214	-133.82060713758\\
215	-134.671247818692\\
216	-133.111651812421\\
217	137.732782484075\\
218	-135.67135584745\\
219	-79.9368733290129\\
220	-138.938211273955\\
221	-77.948645119358\\
222	-40.418611055475\\
223	-55.6316456191429\\
224	161.45186720721\\
225	-59.5976310802337\\
226	-57.2529574918993\\
227	-59.5921012859594\\
228	-61.559803411249\\
229	-63.9345352687952\\
230	-57.6834284866962\\
231	184.691313041033\\
232	135.840634656938\\
233	-64.5492328334983\\
234	-59.1186432173611\\
235	-65.7283790556159\\
236	-71.1106943601566\\
237	-70.9929963164962\\
238	-66.7765867933906\\
239	-71.2088530039196\\
240	-92.0470948617368\\
241	-91.5148651306725\\
242	-95.3031854910638\\
243	-95.1606755258422\\
244	-97.007110058729\\
245	-96.6942418366467\\
246	-96.3584212750132\\
247	-94.6692731571188\\
248	-98.2140663558116\\
249	-96.865571570072\\
250	-95.4877475764941\\
251	-94.0436312133373\\
252	81.8299683254958\\
253	79.894602434914\\
254	-93.7907500271286\\
255	-91.6090538680318\\
256	-84.2642137986522\\
257	73.6253368276298\\
258	-94.2841769743859\\
259	71.3709832183134\\
260	69.5311138275056\\
261	68.0770204294976\\
262	69.7432251114889\\
263	70.921305875706\\
264	71.2124807806389\\
265	73.8872882475067\\
266	72.3127170869797\\
267	69.5234196422594\\
268	-91.64837659112\\
269	-91.267385849786\\
270	-90.0829113837238\\
271	48.7720081775103\\
272	-86.2984999113362\\
273	41.0341590835554\\
274	-84.4004078269217\\
275	-9.61272233736419\\
276	-85.9653552581541\\
277	-46.5129481025672\\
278	6.96834519415335\\
279	-28.1482796866143\\
280	55.9733301256438\\
281	59.694411337704\\
282	-79.9102224833678\\
283	-82.7137558651292\\
284	-83.6669953074631\\
285	-86.0692481193108\\
286	-86.8675541186903\\
287	-87.0300119641267\\
288	-90.6343992005846\\
289	-90.4593328277977\\
290	-90.1308719646472\\
291	-92.5231440844854\\
292	-92.2255499870736\\
293	-95.7557985741872\\
294	-16.9611483711695\\
295	-22.1094307575904\\
296	-93.1098382559703\\
297	-67.5133139862854\\
298	-98.9542889557494\\
299	-97.3698364030132\\
300	-101.844989977089\\
301	-106.322619957765\\
302	-109.603243006378\\
303	-107.258195750878\\
304	-108.697119955193\\
305	-109.999088354437\\
306	-111.038325982191\\
307	-112.739826309207\\
308	-115.160488666417\\
309	-115.33076720651\\
310	-114.408225049202\\
311	-115.328902726842\\
312	-119.036761949368\\
313	-116.077323452576\\
314	-13.8878465654434\\
315	-95.138924880777\\
316	-119.872624499093\\
317	-128.19500461499\\
318	-128.484661246592\\
319	-127.538018582481\\
320	-128.514943198203\\
321	-128.428632774608\\
322	-127.949909767469\\
323	-131.984184116099\\
324	-132.537399143276\\
325	-132.225374261682\\
326	-133.863701786738\\
327	-132.582874793968\\
328	-136.128259976511\\
329	-138.349431530452\\
330	-104.161002717864\\
331	103.762205984758\\
332	-140.3167280092\\
333	70.6644216146686\\
334	122.617006540664\\
335	-141.362996097954\\
336	-59.8509966762195\\
337	162.937339591122\\
338	-53.7976009463297\\
339	-52.4737908533573\\
340	-45.2102178377987\\
341	-23.3173773825662\\
342	11.5248145785473\\
343	-47.5432566094417\\
344	-52.5271434422177\\
345	-67.2965745498921\\
346	-77.0307606834348\\
347	-73.7215970741836\\
};
\addlegendentry{Front Array}

\addplot [color=red, only marks, mark=square, mark size=1pt, mark options={solid, red}]
  table[row sep=crcr]{%
1	-86.0481590397368\\
2	-84.126858074895\\
3	-85.242300757043\\
4	-86.0361502593441\\
5	-86.1986421995612\\
6	-87.086811686267\\
7	-87.6313222114982\\
8	-86.8550297744109\\
9	-80.4314311338612\\
10	-86.6905956144685\\
11	-86.7389740105001\\
12	-85.0893127390608\\
13	-83.0240517489085\\
14	-84.7788458448605\\
15	-77.2915574889918\\
16	-86.3447560452185\\
17	-85.2839448343994\\
18	-84.7089555765855\\
19	-84.2631733999398\\
20	-86.9967479919988\\
21	-85.944492055087\\
22	-85.8218328555578\\
23	-85.8975992360308\\
24	-85.5174212407496\\
25	-86.1783795074193\\
26	-84.0956046455538\\
27	-86.2216286663067\\
28	-83.3390837450804\\
29	-82.8417071183107\\
30	-82.695916218414\\
31	-82.1533110577072\\
32	-82.1650234174278\\
33	-79.1701469916562\\
34	-79.5799853214489\\
35	-81.765712098227\\
36	-80.5686221149955\\
37	-79.5137640274246\\
38	-76.1389742836249\\
39	-76.3903709563954\\
40	-75.9303771996262\\
41	-77.8626650723131\\
42	-78.7703318651693\\
43	-61.6996345804899\\
44	193.614060709532\\
45	182.853038803913\\
46	-76.652532473169\\
47	-76.8627885351795\\
48	-74.6633677117415\\
49	-65.3432762663971\\
50	-68.9442440272358\\
51	-71.2035055546155\\
52	-72.3175506415928\\
53	-72.7321935894352\\
54	-75.6806371352278\\
55	-73.1776305526327\\
56	-75.8221685590894\\
57	-79.1496515516332\\
58	-81.6218770632151\\
59	-81.3260054302295\\
60	-80.5004996726129\\
61	-82.1565751881461\\
62	-82.5573387651572\\
63	-82.7999224394136\\
64	-86.622953560267\\
65	-86.2836845300042\\
66	-87.4842555647567\\
67	-90.3926345174911\\
68	-86.6744186318455\\
69	-90.8002901945021\\
70	-78.420418310848\\
71	-77.8830111059406\\
72	-92.7282317735096\\
73	-93.1502515431354\\
74	-92.0535844996402\\
75	-95.5492514169149\\
76	-96.2144964100968\\
77	-99.9654104281237\\
78	-99.4221975201356\\
79	-96.807328382775\\
80	-101.789416115598\\
81	-101.09938242225\\
82	-102.369922381447\\
83	-105.201419492755\\
84	-105.420724821136\\
85	-108.400557104262\\
86	-107.989090113693\\
87	-109.502942197803\\
88	-109.58038530861\\
89	-112.14610339214\\
90	-114.582787358585\\
91	-117.324301754271\\
92	-118.740146427708\\
93	-118.605681535158\\
94	-120.825936983659\\
95	-121.4555329771\\
96	-122.895892492963\\
97	-125.717488614163\\
98	-125.844529189141\\
99	-126.82063029521\\
100	-111.246329701338\\
101	-127.045267529991\\
102	-121.28716745742\\
103	-67.2784494457417\\
104	127.009138668723\\
105	131.928002496658\\
106	-77.1841039210436\\
107	134.483105126082\\
108	134.939069779826\\
109	-72.9783078465306\\
110	131.347229436827\\
111	112.915093628883\\
112	-67.5754112949158\\
113	140.293928546822\\
114	-67.207579645152\\
115	-67.9980442610462\\
116	141.431361007829\\
117	133.077390711753\\
118	105.501151130384\\
119	135.579807565688\\
120	143.688914908256\\
121	138.294080209085\\
122	143.290676775721\\
123	-86.9051637700668\\
124	-84.9711157868118\\
125	-85.3783341779358\\
126	-84.756153830829\\
127	-84.4541192581174\\
128	-85.751305237716\\
129	-86.0598326298757\\
130	-85.4018933223898\\
131	-88.5281173088384\\
132	-84.3519712809026\\
133	-85.849841751947\\
134	-87.8931288694105\\
135	-88.8023146007982\\
136	-87.3833706759241\\
137	-86.5516914406666\\
138	-86.7699296800834\\
139	-85.5024939281024\\
140	-86.5462909402566\\
141	-86.1233491259763\\
142	-85.3802840410069\\
143	-86.4458659397036\\
144	-86.6120724637988\\
145	-85.5492473472395\\
146	-86.8159361113685\\
147	-85.0319592003075\\
148	-86.687100230717\\
149	-86.3165885913908\\
150	-83.773271382264\\
151	-83.9256409851918\\
152	-81.8955285467484\\
153	-83.6885802324819\\
154	-83.4402997818825\\
155	-82.8777054815614\\
156	-80.538143338656\\
157	-77.9341837598882\\
158	-78.5219274743971\\
159	-80.5376809646045\\
160	-21.1148552563426\\
161	6.92633532047172\\
162	-75.900490289982\\
163	-75.896787550165\\
164	-76.1917506887222\\
165	-74.1100301222134\\
166	-74.7179165202888\\
167	-75.9735628510306\\
168	-76.1545957364989\\
169	-74.5931103928288\\
170	-78.2272081741028\\
171	-78.6697853047516\\
172	-73.3018646215891\\
173	-76.0933200442146\\
174	-78.7006561588622\\
175	-80.5364913738243\\
176	-82.4776926703728\\
177	-81.4287541227502\\
178	-81.3902887914587\\
179	-85.0944514358182\\
180	-87.8984801052409\\
181	-89.5389082925401\\
182	-89.9664698157821\\
183	-92.1580949747455\\
184	-89.9312372180609\\
185	-91.7333076875842\\
186	-92.4195794879506\\
187	-93.1392647851238\\
188	-95.6634220171037\\
189	-99.7627892767753\\
190	-100.213913641966\\
191	-100.22926297808\\
192	-103.575235983849\\
193	-102.928509921753\\
194	-109.953466538999\\
195	-111.066158518114\\
196	-112.709792972981\\
197	-113.639781302157\\
198	-114.689768334181\\
199	-116.127257841494\\
200	-117.783632389781\\
201	-118.621737346655\\
202	-117.263406002322\\
203	-118.210803117552\\
204	-121.924846402549\\
205	-121.305780997987\\
206	-122.383900457451\\
207	-123.393032267562\\
208	-125.021105759521\\
209	-121.063238095364\\
210	-124.575180358529\\
211	-127.678651835698\\
212	-128.905672895513\\
213	-107.077722390322\\
214	-83.5881523345201\\
215	-82.0208547945043\\
216	129.255497157948\\
217	-78.8479936587869\\
218	77.0873590226844\\
219	-78.8477751565114\\
220	125.286573198517\\
221	-75.5619220691074\\
222	-75.3677549211669\\
223	134.103562225899\\
224	131.666665376895\\
225	135.134798305864\\
226	-70.5715249752685\\
227	-67.0255051923449\\
228	98.2478665499656\\
229	136.613287831453\\
230	1.69399233449104\\
231	38.1367635770559\\
232	55.3957783065794\\
233	3.4079084085956\\
234	9.65342660150879\\
235	-6.45267556196994\\
236	11.322860602754\\
237	4.87230588366971\\
238	12.5804710434151\\
239	-6.53007017553194\\
240	-90.8016977745707\\
241	-90.4231843309543\\
242	-18.332386895695\\
243	-86.1341361694072\\
244	-75.4613034131922\\
245	-91.0787350143477\\
246	-95.1591202610868\\
247	-92.67062160745\\
248	-96.9191797387768\\
249	-95.5056542827821\\
250	-94.1155399844424\\
251	-92.683904501077\\
252	-91.7233215360745\\
253	-91.9985723121422\\
254	-93.1663653915461\\
255	-93.3477750856345\\
256	-96.0340806638247\\
257	-95.3240446887038\\
258	-94.0580735590912\\
259	-95.8112608404822\\
260	-96.4682632421741\\
261	-97.3064571128848\\
262	-95.63434762891\\
263	-94.4585027740461\\
264	-94.1552611510693\\
265	-89.3525212945451\\
266	-89.8943793498414\\
267	-91.2227558645173\\
268	-91.8263919173796\\
269	-91.3811061938249\\
270	-87.9889240997595\\
271	-89.2134948648683\\
272	-87.6151571018968\\
273	-84.7985485266109\\
274	-84.9193665452401\\
275	-80.6362761004586\\
276	-84.5489187594887\\
277	-82.0368226140735\\
278	-82.7427576276412\\
279	-4.06537411514414\\
280	-80.3546360845174\\
281	-78.5569306391998\\
282	-78.7087711136878\\
283	-81.3846263324484\\
284	-77.8277263701213\\
285	-84.6886208160903\\
286	-85.4052674791493\\
287	-85.6144857215938\\
288	-89.2811582219192\\
289	-88.9644994950275\\
290	-88.0600357028177\\
291	-91.1360233202141\\
292	-90.8980278790828\\
293	-94.4792561187383\\
294	-96.9915339465337\\
295	-96.902864721601\\
296	-97.7893893407172\\
297	-98.6260253449858\\
298	-90.8110553910578\\
299	-99.3640219115895\\
300	-101.185407537423\\
301	-104.666971583421\\
302	-108.328504293757\\
303	-103.04811725947\\
304	-107.189641636885\\
305	-108.544911997554\\
306	-108.946690713525\\
307	-111.276967628721\\
308	-113.554893307559\\
309	-113.686433706612\\
310	-111.357919063279\\
311	-113.988951614274\\
312	-117.745223280756\\
313	-120.019292133703\\
314	-120.698006033986\\
315	-123.881804367648\\
316	-126.054666058345\\
317	-126.801304531251\\
318	-125.819447730631\\
319	-126.8780949325\\
320	-126.165526599153\\
321	-126.941581621611\\
322	-126.187524309849\\
323	-129.601145055084\\
324	-129.357209561244\\
325	-130.781309418755\\
326	-123.877784319737\\
327	-85.643997741731\\
328	-86.6295217277253\\
329	123.792973326878\\
330	125.219384926956\\
331	125.625031331964\\
332	128.375809480126\\
333	126.926847008779\\
334	129.147630974082\\
335	131.082893599809\\
336	-75.3495932468116\\
337	-77.4298275273921\\
338	-75.6054781656562\\
339	94.6125120763352\\
340	38.4990957047012\\
341	134.938171396911\\
342	137.558938530484\\
343	137.535986901687\\
344	136.878743665406\\
345	137.614596190762\\
346	136.09732299118\\
347	11.8139548739892\\
};
\addlegendentry{Left Array}

\addplot [color=black, dashed, line width=1.2pt, forget plot]
  table[row sep=crcr]{%
0	-300.614060709532\\
0	141.362996097954\\
};

\addplot [color=black, dashed, line width=1.2pt, forget plot]
  table[row sep=crcr]{%
122	-300.614060709532\\
122	141.362996097954\\
};
\addplot [color=black, dashed, line width=1.2pt, forget plot]
  table[row sep=crcr]{%
239	-300.614060709532\\
239	141.362996097954\\
};
\addplot [color=black, dashed, line width=1.2pt, forget plot]
  table[row sep=crcr]{%
347	-300.614060709532\\
347	141.362996097954\\
};
\end{axis}

\end{tikzpicture}%